\documentclass[aps,twocolumn,preprintnumbers,amsmath,amssymb,superscriptaddress,floatfix,nofootinbib,10pt]{revtex4}

\usepackage{graphicx,color,dcolumn,booktabs,bm}
\usepackage{epsfig}
\usepackage{epstopdf}
\usepackage{hyperref}
\usepackage{amsmath}
\allowdisplaybreaks[1]
\usepackage{amsfonts}
\usepackage{amssymb}
\usepackage{multirow}
\usepackage{makecell}
\usepackage[dvipsnames]{xcolor}
\usepackage{ulem}
\usepackage{longtable}
\usepackage{array}
\usepackage{subfigure}

\def\lrpartial{\buildrel\leftrightarrow\over\partial}
\begin{document}

\title{Proposal to detect a moving triangle singularity in $\psi(2S) \to \pi^+ \pi^- K^+ K^-$ process}

\author{Qi Huang}\email{huangqi@ucas.ac.cn}
\affiliation{School of Physical Sciences, University of Chinese Academy of Sciences (UCAS), Beijing 100049, China}
\author{Jia-Jun Wu} \email{wujiajun@ucas.ac.cn}
\affiliation{School of Physical Sciences, University of Chinese Academy of Sciences (UCAS), Beijing 100049, China}

\date{\today}

\begin{abstract}

In this work, we propose that there exists a moving triangle singularity in the $\psi(2S) \to \pi^+ \pi^- K^+ K^-$ process, whose position can vary from 1.158 to 1.181 GeV in the invariant mass spectrum of $K^+K^-$. After a precise analysis on this process, it turns out that after doing some cuts on $m_{\pi^+\pi^-}$, experiments do have the opportunity to observe this triangle singularity. In addition, when changing the cuts on $m_{\pi^+\pi^-}$, the movement of the predicted triangle singularity can also be observed. Thus, we suggest future experiments, especially Super Tau-Charm Facility (STCF), to do an anlysis on the $\psi(2S) \to \pi^+ \pi^- K^+ K^-$ process to verify our prediction.

\end{abstract}


\maketitle

\section{Introduction}\label{sec1}

In the past decades, the triangle singularity that proposed by L. D. Landau in 1959~\cite{Landau:1959fi} has been recognized to play important roles in understanding a series of anomalous experimental observations. 
For example, after introducing the triangle loop composed by kaons, Refs.~\cite{Wu:2011yx,Aceti:2012dj,Wu:2012pg, Achasov:2015uua, Du:2019idk} successfully explained the isospin breaking process $\eta(1405) \to \pi^0 f_0(980)$ \cite{BESIII:2012aa} and Ref.~\cite{Ketzer:2015tqa} interpreted the nature of $a_1(1420)$ through $\pi p \to a_1(1260) \to f_0(980) \pi$ process. 
Especially, in recent years, with the discoveries of a series of exotic states such as $Z_c$~\cite{Ablikim:2013mio,Liu:2013dau,Xiao:2013iha,Ablikim:2013wzq,Ablikim:2013xfr,Ablikim:2013emm,Ablikim:2017oaf}, $P_c$~\cite{Aaij:2015tga,Aaij:2019vzc} and X(2900)~\cite{Aaij:2020hon,Aaij:2020ypa}, many researches on triangle singularity have been carried out~\cite{Wu:2011yx, Aceti:2012dj, Wu:2012pg, Ketzer:2015tqa, Wang:2013cya,Wang:2013hga, Achasov:2015uua, Liu:2015taa,Liu:2015fea,Guo:2015umn,Szczepaniak:2015eza, Guo:2016bkl, Bayar:2016ftu, Wang:2016dtb, Pilloni:2016obd, Xie:2016lvs, Szczepaniak:2015hya, Roca:2017bvy,
Debastiani:2017dlz, Samart:2017scf, Sakai:2017hpg, Pavao:2017kcr, Xie:2017mbe, Bayar:2017svj,Liang:2017ijf, Oset:2018zgc, Dai:2018hqb, Dai:2018rra, Guo:2019qcn, Liang:2019jtr, Nakamura:2019emd,Liu:2019dqc, Jing:2019cbw, Braaten:2019gfj, Sakai:2020ucu, Sakai:2020fjh, Molina:2020kyu, Braaten:2020iye, Alexeev:2020lvq, Ortega:2020ayw,Shen:2020gpw,Du:2019idk,Liu:2020orv,Achasov:2019wvw} (for a recent review, see Ref.~\cite{Guo:2019twa}), which imply that these exotic states can be related to some specific triangle singularities.

However, as pointed out by Ref. \cite{Huang:2020kxf}, although triangle singularity can successfully explain so many experimental phenomena, until now the anomaly peak structure due to the triangle singularity has not been fully confirmed by any experiments. Especially, recently, we noticed that the COMPASS experiment reanalyzed the $\pi p \to a_1(1260) \to f_0(980) \pi \to 3\pi$ process \cite{COMPASS:2020yhb}. By using the triangle singularity produced by the kaon loop, they successfully explained the peak of $a_1(1420)$ without introducing the Breit-Wigner structure. Although it obviously shows the importance of triangle singularity in the hadron reaction, as pointed out by Ref. \cite{COMPASS:2020yhb}, the fit where triangle singularity participates in is just slightly better than the Breit-Wigner model, which indicates that the exsiting data still can not rule out the Breit-Wigner model for $a_1(1420)$. Thus, in our view, we still need further evidence for the triangle singularity.

According to the conclusions of our previous work \cite{Huang:2020kxf}, there are several difficulties to search a perfect process which can show the triangle singularity phenomena.
For example, the interference from the thresholds cusp will make the distinction of triangle singularity difficult, and the unknown vertices in the triangle loop of most processes will make the precise prediction impossible on the theoretical side. Thus, Ref. \cite{Huang:2020kxf} proposed that it is very possible for experiments to detect a pure triangle singularity in the $\psi(2S) \to p\bar{p}\eta$ process, where the triangle loop is composed by $J/\psi$, $\eta$ and proton. 
Under this situation, the position of the triangle singularity is 80 MeV above the $J/\psi\eta$ threshold. Also, since $J/\psi$, $\eta$, and the proton are all very narrow particles, the signal of the triangle singularity is very sharp, which can be distinguished from excited nucleons easily. 
In addition, $\psi(2S) \to J/\psi \eta$, $J/\psi \to p\bar{p}$ and $p\eta \to p\eta$ processes can all be constrained by the experimental data. 
As a result, Ref. \cite{Huang:2020kxf} do find a triangle singularity at the right shoulder of $N(1535)$, whose width is about 5 MeV, and it may be observed by the future experiments such as Beijing Spectrometer (BESIII) and STCF.

\begin{figure}[tbp]
	\centering
    \includegraphics[width=0.9\linewidth]{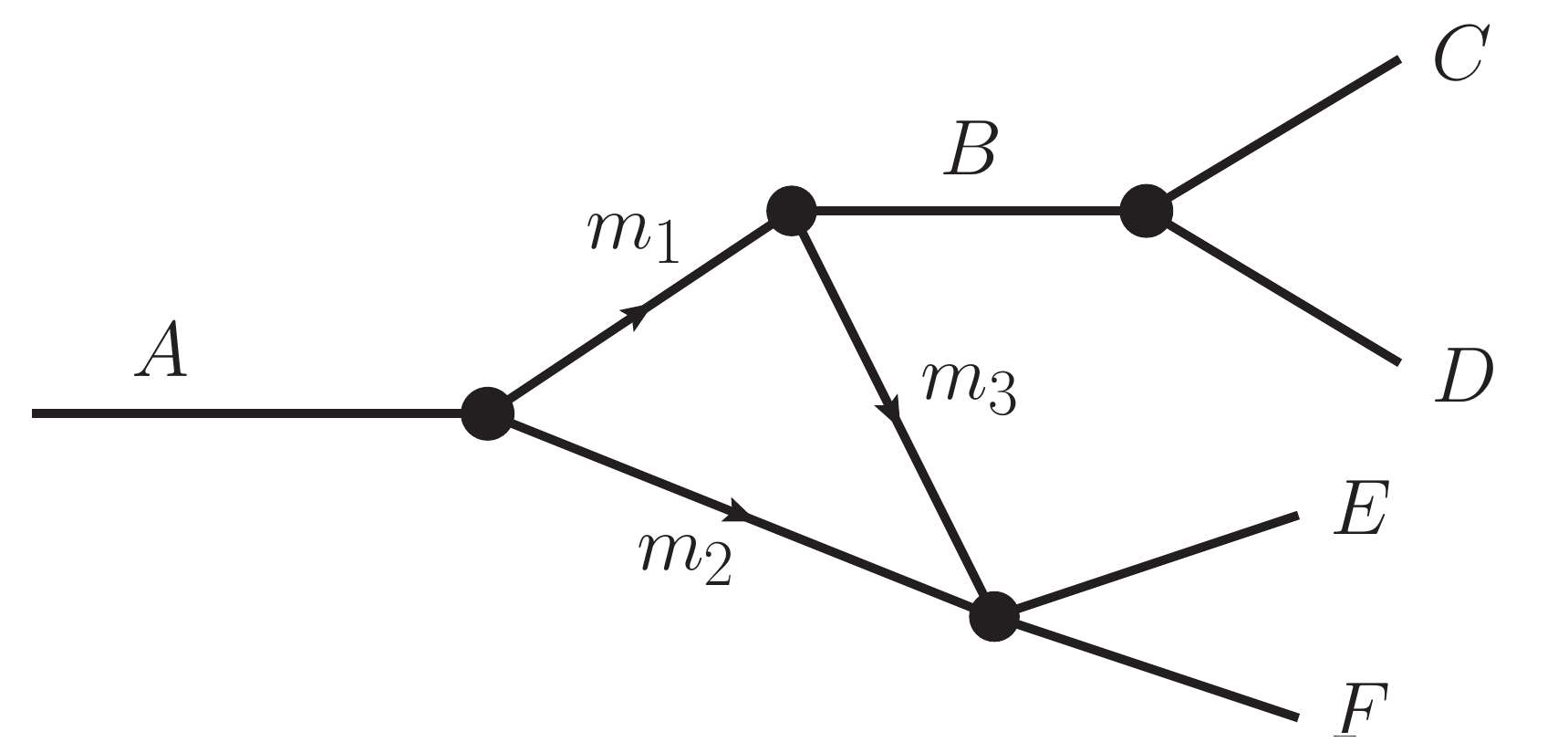}
\caption{Kinematical mechanism of the production of a moving triangle singularity.}
\label{fig:loop}
\end{figure}

Usually, in most papers the triangle singularity effects are studied in a process with three-body final state. However, we find that in the process with four-body final states, there will exist very interesting phenomena due to the triangle singularity.
In Fig. \ref{fig:loop}, we show the diagram for a $1\to 4$ process, where two final particles labeled as $C$ and $D$ are from the decay of particle $B$.
Then for the fixed masses of mother particle $A$, intermediate particles $1$, $2$, and $3$, the relationship between the invariant masses of particles $(C, D)$ and $(E, F)$ can be derived by the kinematic condition of triangle singularity known as Coleman-Norton theorem \cite{Shen:2020gpw,Huang:2020kxf},
\begin{eqnarray}
  \sqrt{(\omega_{EF}-\omega_{2})^2-m_3^2}&=&q_{EF}-q_{2}, \label{eq:cond1}\\
  \frac{q_{EF}-q_{2}}{\omega_{EF}-\omega_2}&>&\frac{q_{2}}{\omega_{2}},
  \label{eq:cond2}
\end{eqnarray}
where
\begin{eqnarray}
  \omega_{EF} &=& \frac{m_A^2+m_{EF}^2-m^2_{CD}}{2m^2_A},\\
  \omega_{2} &=&\frac{m_A^2+m_{2}^2-m^2_{1}}{2m^2_A},\\
  q_{EF}&=& \sqrt{\omega_{EF}^2-m_{EF}^2},\\
  q_{2}&=& \sqrt{\omega_{2}^2-m_{2}^2},
\end{eqnarray}   
and $m_{CD/EF}$ is the invariant masses of particles $(C, D)/(E, F)$, respectively.
In principle, if the triangle singularity is permitted in this process, there would be a series of $(m_{CD}, m_{EF})$ that can satisfy the above equations.  
Thus, if we fix the value of $m_{CD}$ in the permitted kinematical range, due to the triangle singularity there will be a peak structure in the invariant mass spectrum of $(E,F)$ and the peak position can be solved exactly from Eqs. (\ref{eq:cond1}) and (\ref{eq:cond2}).
Then once the value of $m_{CD}$ is changed, such peak will also move, i.e., the peak position solved by the above equations will be changed.
Here we call it as "a moving triangle singularity", which should be an interesting phenomena for both theorists and experimentalists.

Thus, as the first step, in this work we propose that a moving triangle singularity can really happen in the $\psi(2S) \to \pi^+ \pi^- K^+ K^-$ process. As shown in Fig. \ref{fig:loop-and-background} (a), $\psi(2S)$ decays into $J/\psi$ and $\eta$ first, then $J/\psi$ decays into $\rho^0$ and $\pi^0$, after that $\rho^0$ decays into $\pi^+\pi^-$ and a re-scattering happens between $\eta$ and $\pi^0$ and transit into $K^+ K^-$. 
Obviously, the $\pi^0$ decay from the $J/\psi$ has large velocity and it can catch $\eta$ easily, which causes the triangle singularity.
On the other hand, since the the width of $\rho$ is very large, i.e., around $140$ MeV, we expect that the triangle singularity produced can move in a considerable range.
From Eqs.(\ref{eq:cond1},\ref{eq:cond2}), 
we find that the position of the triangle singularity produced by this process can vary from 1.158 GeV to 1.181 GeV in the invariant mass spectrum of $K^+K^-$, i.e., there exists about $23$ MeV kinematic space for the triangle singularity to move.
Thus, in the current paper, we will do a detailed analysis on this triangle singularity and explore the possibility if future experiments can verify our predictions.

This paper is organized as follows. After the introduction, we give the main decay mechanisms of $\psi(2S) \to \pi^+ \pi^- K^+ K^-$ process in Sec. \ref{sec:2-mechanism}. Then the numerical results and corresponding discussions are given in Sec. \ref{sec3}. Finally, a summary is presented.

\section{Main decay mechanisms of $\psi(2S) \to \pi^+ \pi^- K^+ K^-$ process}\label{sec:2-mechanism}

\begin{figure}[htbp]
  \centering
  \includegraphics[width=0.45\linewidth]{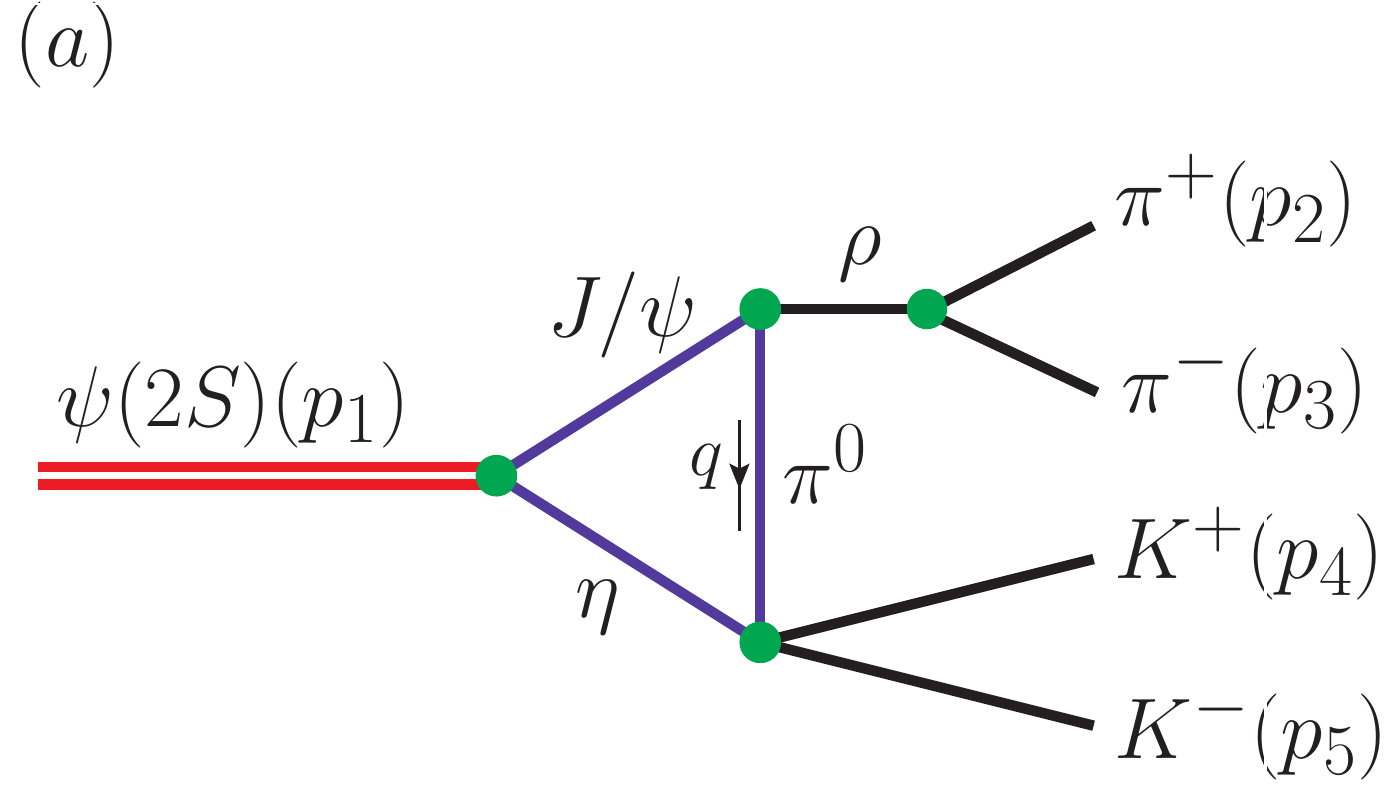} \
  \includegraphics[width=0.45\linewidth]{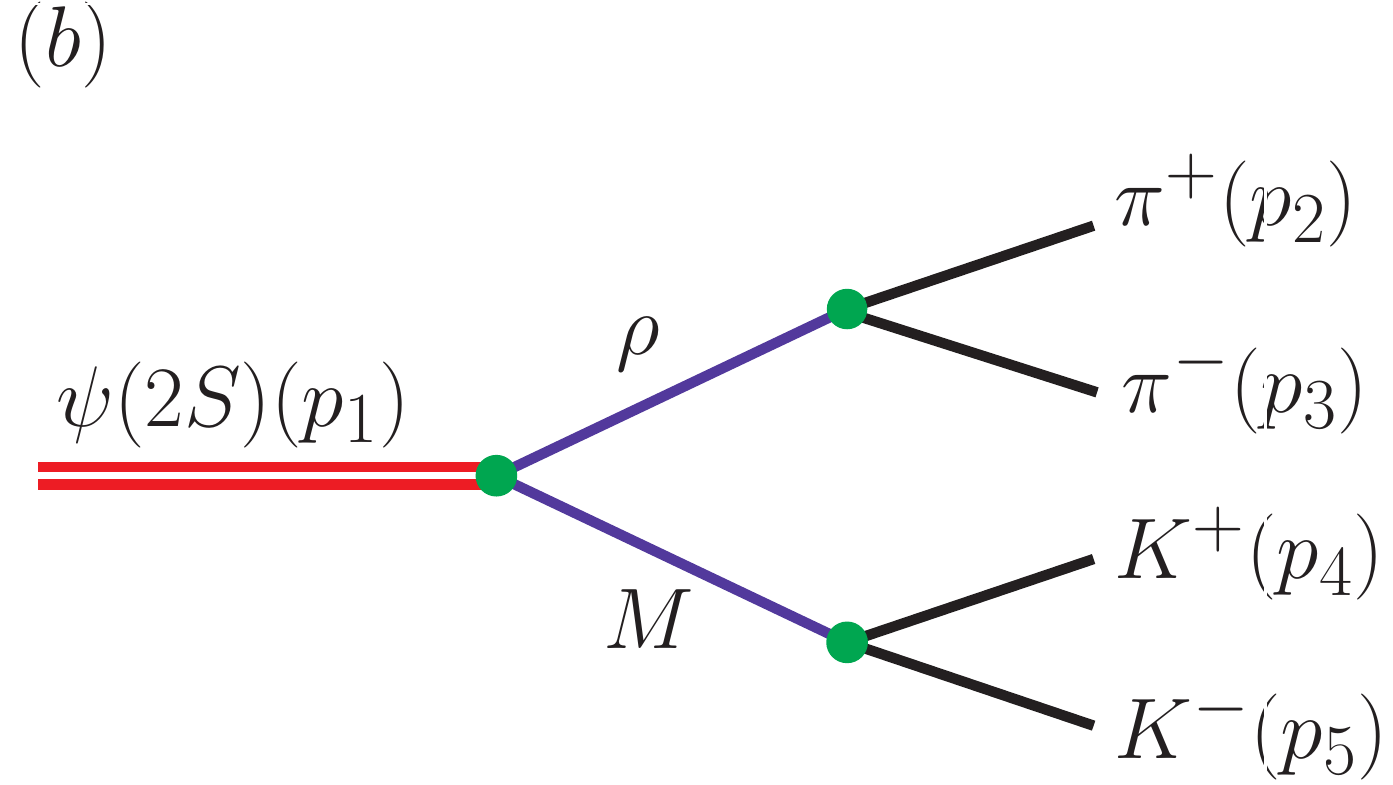}
\caption{The Feynman diagrams describing the process $\psi(2S) \to p \bar{p} \eta / p \bar{p} \pi$. (a): loop diagram where triangle singularity happens; (b): tree diagram called "background".}
\label{fig:loop-and-background}
\end{figure}

The typical diagrams for the $\psi(2S) \to \pi^+ \pi^- K^+ K^-$ process are given in Fig.\ref{fig:loop-and-background}. Here, Fig.\ref{fig:loop-and-background}(a) presents the triangle loop diagram for $\psi(2S) \to \pi^+ \pi^- K^+ K^-$, which is similar to our previous work on $\psi(2S) \to p \bar{p} \eta / p \bar{p} \pi^0$ process~\cite{Huang:2020kxf}.
On the other hand, the corresponding tree diagram considered as the "background" is shown in Fig.\ref{fig:loop-and-background} (b), where $M$ denotes an intermediate meson. 

It is clear that the diagram shown in Fig. \ref{fig:loop-and-background} (a), which is similar as Fig.~\ref{fig:loop}, is a nice place to study the moving triangle singularity. 
In the triangle loop diagram, $\psi(2S)$ decays into $J/\psi$ and $\eta$ first, 
then $J/\psi$ decays into $\rho^0$ and $\pi^0$.
When $\pi^0$ moves in the same direction as $\eta$ and catches up with it, it scatters to the charged kaon pair and the triangle singularity happens.
Obviously, the value of the invariant mass of kaon pair at the triangle singularity point are determined by the relative velocity between $\pi^0$ and $\eta$.
Actually, the velocity of the $\pi^0$ emitted by $J/\psi$ will change because of the broad $\rho$ meson.
As a result, the peak position due to the triangle singularity in the invariant mass of kaon pair should move with the invariant mass of $\pi^+ \pi^-$. 
By applying Eqs.(\ref{eq:cond1},\ref{eq:cond2}), we find that when the invariant mass of $\pi^+ \pi^-$ changes within $[m_\rho - \Gamma_\rho,m_\rho+\Gamma_\rho]$, where $m_\rho \sim 770$ MeV and $\Gamma_\rho \sim 140$ MeV denote the mass and width of $\rho$ meson, respectively, the position due to the triangle singularity at the invariant mass spectrum of final koan pair changes in the interval $[1.158~\mathrm{GeV},1.181~\mathrm{GeV}]$. 
In this energy range, the main "background" should comes from $M=a_0(980)$ and $a_2(1320)$ as shown in Fig.\ref{fig:loop-and-background} (b).
Furthermore, we will give an estimation of the other possible background in the next section.

The tree diagram in Fig. \ref{fig:loop-and-background} gives the dominant contribution to $\psi(2S) \to \pi^+ \pi^- K^+ K^-$ process, thus, our target is to find a way to make the signal caused by the triangle singularity visible. 
In this work, we adopt the effective Lagrangian approach to do the calculation, and the general forms of the relevant effective Lagrangians are constructed as
\begin{eqnarray}
  \mathcal{L}_{\mathcal{V}\mathcal{V}\mathcal{P}} &=& g_{\mathcal{V}\mathcal{V}\mathcal{P}} \varepsilon^{\mu\nu\alpha\beta} \partial_\mu \mathcal{V}_\nu \partial_\alpha \mathcal{V}_\beta \mathcal{P},\\
  \mathcal{L}_{\mathcal{V}\mathcal{P}\mathcal{P}} &=& i g_{\mathcal{V}\mathcal{P}\mathcal{P}} \mathcal{V}^\mu \mathcal{P} \lrpartial_\mu \mathcal{P},\\
  \mathcal{L}_{\mathcal{S}\mathcal{P}\mathcal{P}} &=& g_{\mathcal{S}\mathcal{P}\mathcal{P}}\mathcal{S}\mathcal{P}\mathcal{P},\\
  \mathcal{L}_{\mathcal{V}\mathcal{V}\mathcal{S}} &=& g_{\mathcal{V}\mathcal{V}\mathcal{S}} \mathcal{V}^\mu \mathcal{V}_\mu \mathcal{S},\\
  \mathcal{L}_{\mathcal{V}\mathcal{V}\mathcal{T}} &=& g_{\mathcal{V}\mathcal{V}\mathcal{T}} \mathcal{V}^\mu \mathcal{V}^\nu \mathcal{T}_{\mu\nu},\\
  \mathcal{L}_{\mathcal{T}\mathcal{P}\mathcal{P}} &=& g_{\mathcal{T}\mathcal{P}\mathcal{P}} \mathcal{T}^{\mu\nu} \partial_\mu \mathcal{P} \partial_\nu \mathcal{P},
\end{eqnarray}
where $\mathcal{P}$, $\mathcal{S}$, $\mathcal{V}$, and $\mathcal{T}$ are the fields of pesudo-scalar, scalar, vector, and tensor mesons, respectively.

Then, the amplitudes of the triangle loop diagram and tree diagram given in Fig. \ref{fig:loop-and-background} can be obtained straightforwardly. 
For the triangle loop diagram, we can get that
\begin{eqnarray}
  &&\mathcal{M}^{\mathrm{Loop}}\nonumber\\
  && \quad = i \int \frac{d^4 q}{(2\pi)^4} \frac{\mathcal{F}(p_2+p_3+q,m_{J/\psi},\Lambda_{J/\psi})}{(p_2+p_3+q)^2-m_{J/\psi}^2+im_{J/\psi}\Gamma_{J/\psi}}\nonumber\\
  && \qquad \times \frac{\mathcal{F}(q,m_{\pi^0},\Lambda_{\pi^0})}{q^2-m_{\pi^0}^2+im_{\pi^0}\Gamma_{\pi^0}}\nonumber \frac{\mathcal{G}_{\omega\xi}(p_2+p_3,m_{\rho})}{(p_2+p_3)^2 - m_\rho^2 + im_\rho\Gamma_\rho}\\
  && \qquad \times \frac{\mathcal{F}(p_4+p_5-q,m_\eta,\Lambda_\eta)}{(p_4+p_5-q)^2-m_{\eta}^2+im_{\eta}\Gamma_{\eta}} \mathcal{M}_{\eta\pi^0 \to K^+ K^-}\nonumber\\
  && \qquad \times g_{\psi(2S) J/\psi \eta} \varepsilon^{\mu\nu\alpha\beta} p_{1\mu} \epsilon_{\psi(2S)\nu} (p_{2\alpha}+p_{3\alpha}+q_\alpha) \nonumber\\
  && \qquad \times g_{J/\psi \rho \pi^0} \varepsilon^{\lambda\tau\chi\omega} (p_{2\lambda}+p_{3\lambda}+q_\lambda) (p_{2\chi}+p_{3\chi})\nonumber\\
  && \qquad \times g_{\rho \pi \pi} (p_3^\xi - p_2^\xi) \mathcal{G}_{\beta\tau}(p_2+p_3+q,m_{J/\psi}),\label{eq:M-Loop}
\end{eqnarray}
where $\mathcal{G}_{\mu\nu}(p,m) = -g_{\mu\nu} + \frac{p_\mu p_\nu}{m^2}$ is the projection operator, 
$\mathcal{M}_{\eta \pi^0 \to K^+ K^-}$ is the amplitude of $\eta \pi^0 \to K^+ K^-$ transition, which is described with the chiral unitary approach \cite{Gasser:1983yg,Bernard:1995dp,Oller:1997ti,Oller:1998hw,Kaiser:1998fi,Locher:1997gr,Nieves:1999bx,Pelaez:2006nj,Xie:2014tma,Liang:2014tia,Toledo:2020zxj,Ikeno:2021kzf,Molina:2019udw} and the detailed expressions are derived in the appendix, 
and $\mathcal{F}(q,m,\Lambda) = \frac{\Lambda^4}{(q^2-m^2)^2+\Lambda^4}$ is the form factor, which is used to describe the structure effects of interaction vertices and off-shell effects of internal particles, also, the introduction of this form factor will help us avoid the ultraviolet divergence. 
We want to note here that when the triangle singularity happens, all the internal particles are on-shell. At that time, we have $\mathcal{F}(p,m,\Lambda) =1$, i.e., these form factors will not affect the strength of triangle singularity at the peak position.  
In the current calculation, as done in Ref. \cite{Huang:2020kxf}, $\Lambda_{J/\psi,\eta,\pi^0}$ is set as $ m_{J/\psi,\eta,\pi^0} + \alpha \Lambda_{\mathrm{QCD}}$, where $\alpha$ is a free parameter and $\Lambda_{\mathrm{QCD}} = 0.22~\mathrm{GeV}$.
The value of $\alpha$ is taken as $1$ because the change of this $\alpha$ affects little the behavior of $\mathcal{M}^{\mathrm{Loop}}$.
A similar conclusion is proven by the numerical calculation in the previous work on the $\psi(2S) \to p\bar{p}\eta/p\bar{p}\pi^0$ process \cite{Huang:2020kxf}.

Next, for the tree diagram, i.e., Fig. \ref{fig:loop-and-background} (b), when the intermediate meson $M$ is $a_0(980)$, considering that $a_0(980)$ is very close to the $K^+ K^-$ threshold, the propagator of $a_0((980)$ should be expressed in Flatte form \cite{Wu:2008hx}. 
Thus, this amplitude can be written as
\begin{eqnarray}
  &&\mathcal{M}^{\mathrm{Tree}}_{a_0(980)}\nonumber\\ 
  && \quad \quad = g_{\psi(2S) a_0(980)^0 \rho^0} g_{\rho^0\pi^+\pi^-}\epsilon_{\psi(2S)\mu}  (p_{3\nu} - p_{2\nu})\nonumber\\
  && \quad \qquad \times \frac{g_{a_0(980)^0 K^+ K^-} }{m_{a_0(980)^0}^2 - m_{45}^2 - i m_{45} \Gamma_{a_0(980)^0}(m_{45})}\nonumber\\
  && \qquad \quad \times \frac{\mathcal{G}^{\mu\nu}(p_2+p_3,m_{\rho^0})}{m_{23}^2-m_{\rho^0}^2+im_{\rho^0}\Gamma_{\rho^0}},
\end{eqnarray}
with
\begin{eqnarray}
  &&\Gamma_{a_0(980)^0}(m_{45}) = \Gamma_{a_0(980)^0}^{\eta \pi^0}(m_{45}) + \Gamma_{a_0(980)^0}^{K^+ K^-}(m_{45}),\\
  &&\Gamma_a^{bc} (s) = \frac{g_{abc}^2}{16\pi\sqrt{s}} \rho_{bc}(s),\\
  &&\rho_{bc}(s) = \frac{\sqrt{(s-(m_b-m_c)^2)(s-(m_b+m_c)^2)}}{s},
\end{eqnarray}
where $m_{45}$ and $m_{23}$ are the invariant masses of $K^+K^-$ and $\pi^+\pi^-$ respectively, $g_{abc}$ is the coupling constant. As done in Ref. \cite{Wu:2008hx}, in this work, we set $g_{a_0(980)^0 K^+ K^-} = 2.54$ GeV and $g_{a_0(980)^0 \pi^0\eta} = 3.33$ GeV.

Then, for the tree diagram where the intermediate meson $M$ is $a_2(1320)$, the amplitude can be written as
\begin{eqnarray}
  &&\mathcal{M}^{\mathrm{Tree}}_{a_2(1320)}\nonumber\\
  && \qquad = g_{\psi(2S)a_2(1320)^0\rho^0} g_{\rho^0\pi^+\pi^-}\epsilon_{\psi(2S)}^\mu (p_{2\xi}-p_{3\xi}) p_{4\alpha}p_{5\beta}\nonumber\\
  && \qquad \quad \times \frac{g_{a_2(1320)^0K^+K^-} \mathcal{G}_{\mu\nu\alpha\beta}(p_4+p_5,m_{a_2(1320)^0})}{m_{45}^2-m_{a_2(1320)^0}^2+im_{a_2(1320)^0}\Gamma_{a_2(1320)^0}}\nonumber\\
  && \qquad \quad \times \frac{\mathcal{G}^{\nu\xi}(p_2+p_3,m_{\rho^0})}{m_{23}^2-m_{\rho^0}^2+im_{\rho^0}\Gamma_{\rho^0}},
\end{eqnarray}
where
\begin{eqnarray}
  &&\mathcal{G}^{\mu\nu\alpha\beta}(p,m)\nonumber\\
  &&\qquad = \frac{1}{2} \left( \mathcal{G}^{\mu\alpha}(p,m)\mathcal{G}^{\nu\beta}(p,m) + \mathcal{G}^{\mu\beta}(p,m)\mathcal{G}^{\nu\alpha}(p,m) \right)\nonumber\\
  &&\qquad \quad - \frac{1}{3} \mathcal{G}^{\mu\nu}(p,m) \mathcal{G}^{\alpha\beta}(p,m).
\end{eqnarray}

Finally, the differential decay width of the $\psi(2S) \to \pi^+ \pi^- K^+ K^-$ process can be expressed as \cite{Jing:2020tth}
\begin{eqnarray}
  d\Gamma &=& \sum|\mathcal{M}^{\mathrm{Tree}}_{a_0(980)} + \mathcal{M}^{\mathrm{Tree}}_{a_2(1320)} + \mathcal{M}^{\mathrm{Loop}}|^2\nonumber\\
  && \times \frac{m_{234} m_{45} |\vec{p}_{\pi}|^\ast}{48(2\pi)^6 m_{\psi(2S)}^3} dm_{234} dm_{45} dm_{23} d\Omega^\ast_\pi,
\end{eqnarray}
where $\sum$ denotes the summation and average over the spin of $\psi(2S)$, $|\vec{p}_{\pi}|^\ast$ and $\Omega^\ast_\pi$ are the modulus of the 3-momentum and solid angle of $\pi^+/\pi^-$ in the $\rho^0$ rest frame respectively.

\section{Numerical results and discussions}\label{sec3}

\subsection{Determining the coupling constants}

Before presenting our numerical results, we need to determine the relevant coupling constants needed first. 
For the coupling constants $g_{\psi(2S)J/\psi\eta}$, $g_{J/\psi \rho^0 \pi^0}$, $g_{\rho^0\pi^+\pi^-}$, $g_{a_2(1320)^0K^+K^-}$ and $g_{\psi(2S)a_2(1320)^0\rho^0}$, since experiments have measured the branching ratios of the corresponding processes \cite{Zyla:2020zbs}, these coupling constants can be extracted from the data given in RPP \cite{Zyla:2020zbs}, both of which are collected in Table \ref{tab:coupling-constants}.

\begin{table}[htpb]
	\renewcommand\arraystretch{1.5}
  \centering
	\caption{The values of relevant coupling constants extracted from the corresponding branching ratios \cite{Zyla:2020zbs}.}
	\label{tab:coupling-constants}
  \begin{tabular}{ccc}
  \toprule[1pt]
  Coupling constant & Branching ratio \cite{Zyla:2020zbs} & Value\\
  \midrule[1pt]
  $g_{\psi(2S)J/\psi\eta}$ & $(3.37 \pm 0.05) \times 10^{-2}$ & (0.218 $\pm$ 0.003)\\ 
  $g_{J/\psi \rho^0 \pi^0}$ & $(5.6 \pm 0.7) \times 10^{-3}$ &  $(2.535 \pm 0.159) \times 10^{-3}$\\ 
  $g_{\rho^0 \pi^+ \pi^-}$ & $\sim 100\%$ & $\sim$ 7.242\\
  $g_{a_2(1320)^0K^+K^-}$ & $ (2.45 \pm 0.8) \times 10^{-2}$ &  (5.669 $\pm$ 0.956)\\ 
  $g_{\psi(2S)a_2(1320)^0\rho^0}$ & $\sim 8.67 \times 10^{-5}$ & $\sim 6.215 \times 10^{-4}$\\ 
  \bottomrule[1pt]
  \end{tabular}
\end{table}

For the $g_{\psi(2S)a_0(980)\rho^0}$, since the branching ratio of $\psi(2S) \to a_0(980) \rho$ process is still absent in RPP \cite{Zyla:2020zbs}, by assuming that $J/\psi$ and $\psi(2S)$ have very similar properties, we naively estimate the coupling constants between $\psi(2S)$, $a_0(980)$, and $\rho$ by the following way. 
From RPP, we can get that the branching ratios of $J/\psi \to f_0(980) \omega$, $J/\psi \to f_0(980) \phi$, $J/\psi \to f_0(980) \phi \to \phi \pi^+ \pi^-$, and $\psi(2S) \to f_0(980) \phi \to \phi \pi^+ \pi^-$ processes are $(1.4 \pm 0.5) \times 10^{-4}$, $(3.2 \pm 0.9) \times 10^{-4}$, $(2.59 \pm 0.34) \times 10^{-4}$, and $(7.5 \pm 3.3) \times 10^{-5}$, respectively \cite{Zyla:2020zbs}. 
Thus the branching ratio of $\psi(2S) \to a_0(980) \rho$ process might be estimated as
\begin{eqnarray}
  &&\mathcal{B}(\psi(2S) \to a_0(980)^0 \rho^0) \nonumber\\
  &&\quad \approx \mathcal{B}(\psi(2S) \to f_0(980)\omega) \nonumber\\
  &&\quad = \frac{\mathcal{B}(J/\psi \to f_0(980)\omega)}{\mathcal{B}(J/\psi \to f_0(980)\phi)} \mathcal{B}(\psi(2S) \to f_0(980)\phi) \nonumber\\
  &&\quad = \frac{\mathcal{B}(J/\psi \to f_0(980)\omega)}{\mathcal{B}(J/\psi \to f_0(980)\phi)} \mathcal{B}(J/\psi \to f_0(980)\phi) \nonumber\\
  &&\quad\times \frac{\mathcal{B}(\psi(2S) \to f_0(980)\phi \to \phi \pi^+ \pi^-)}{\mathcal{B}(J/\psi \to f_0(980)\phi \to \phi \pi^+ \pi^+)} \nonumber\\
  &&\quad = \mathcal{B}(J/\psi \to f_0(980)\omega) \frac{\mathcal{B}(\psi(2S) \to f_0(980)\phi \to \phi \pi^+ \pi^-)}{\mathcal{B}(J/\psi \to f_0(980)\phi \to \phi \pi^+ \pi^+)} \nonumber\\
  &&\quad = (4.054 \pm 2.358) \times 10^{-5}.
\end{eqnarray}
Then the coupling constant for the vertex of $\psi(2S)\to a_0(980)^0\rho^0$ is calculated as $g_{\psi(2S)a_0(980)^0\rho^0} = (9.6 \pm 3.1) \times 10^{-4}$ GeV.

\subsection{The signal of triangle singularity} \label{sec:3-Loop}

\begin{figure}[htbp]
  \centering
  \includegraphics[width=1.0\linewidth]{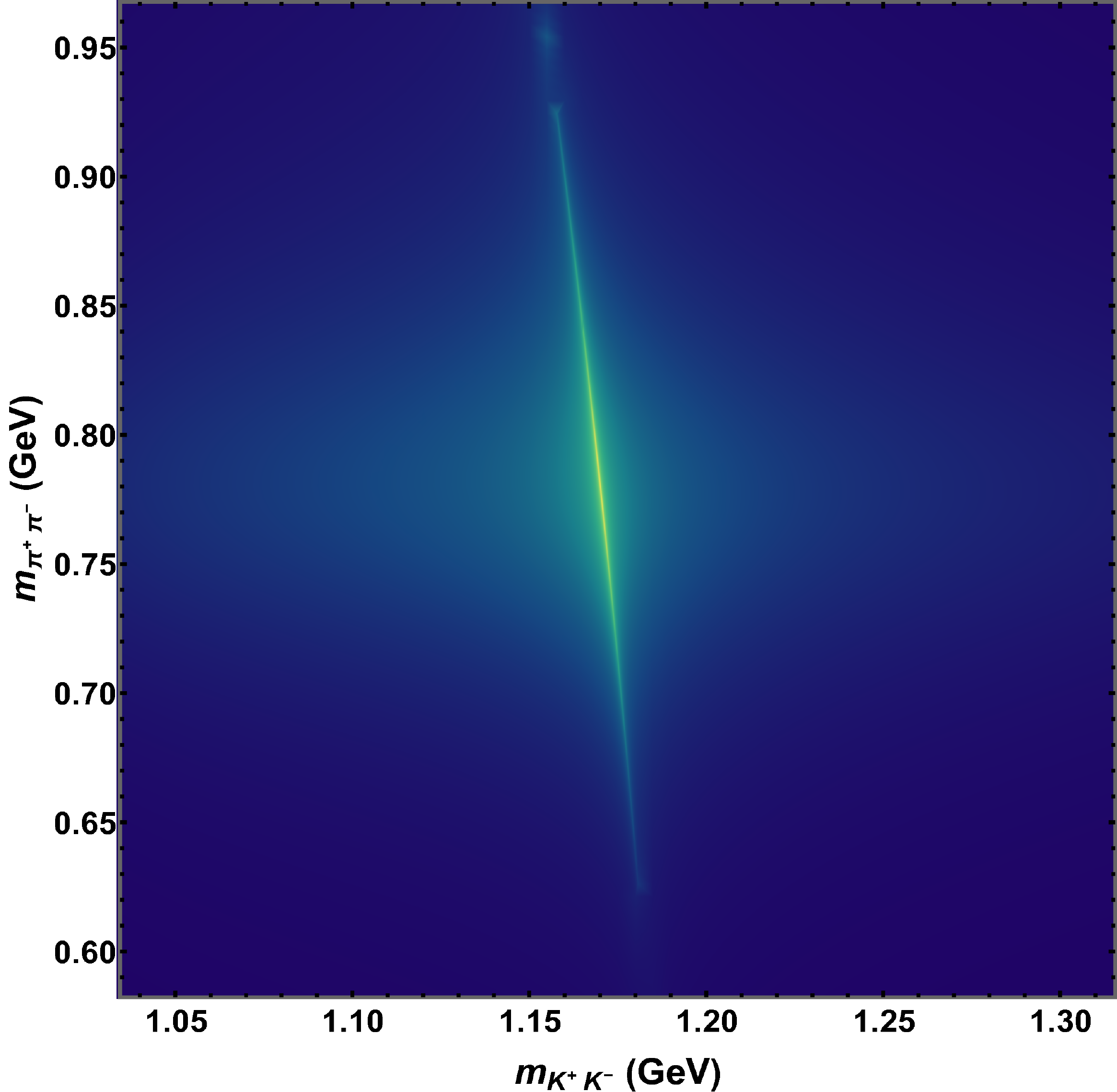}
\caption{The Dalitz plot of the $\psi(2S) \to \pi^+ \pi^- K^+ K^-$ process after considering the contribution of Fig. \ref{fig:loop-and-background} ($a$) only, where $m_{\pi^+\pi^-}$ and $m_{K^+K^-}$ are limited in [0.59 GeV, 0.96 GeV] and [1.04 GeV, 1.31 GeV] respectively, and the thin band in the middle is the contribution of the triangle singularity.}
\label{fig:loop-dalitz}
\end{figure}

With all the preparations above, now we can present our numerical results. 
First of all, we want to show the behavior of the loop diagram of $\psi(2S) \to \pi^+ \pi^- K^+ K^-$ process. 
In Fig. \ref{fig:loop-dalitz}, the Dalitz plot of the loop diagram only is presented, where a thin band totally caused by the triangle singularity can be clearly seen. 
In this figure, $m_{\pi^+\pi^-}$ and $m_{K^+K^-}$ are limited in [0.59 GeV, 0.96 GeV] and [1.04 GeV, 1.31 GeV] respectively. 
Here, the interval of $m_{\pi^+\pi^-}$ just covers the $\rho$ meson as $[m_\rho-1.3\Gamma_\rho, m_\rho+1.3\Gamma_\rho]$, and the range of $m_{K^+K^-}$ is just around the range where the triangle singularity is happening. 
From Fig. \ref{fig:loop-dalitz}, it is clear that when $m_{\pi^+ \pi^-}$ is closer to the mass of $\rho$ meson, the brightness of the thin band is higher.  
In addition, when $m_{\pi^+ \pi^-}$ is smaller than $m_\rho-\Gamma_\rho$ (0.62 GeV) or larger than $m_\rho+\Gamma_\rho$ (0.92 GeV), the color of the thin band is too gloomy to be distinguished from the region where the triangle singularity doesn't happen, i.e., the strength of the triangle singularity is suppressed tremendously by the propagator of $\rho$ meson when the difference between $m_{\pi^+\pi^-}$ and $m_\rho$ is larger than $\Gamma_\rho$. 
Thus, in the loop diagram of $\psi(2S) \to \pi^+ \pi^- K^+ K^-$ process, when $m_{\pi^+\pi^-}$ changes in the interval [0.59 GeV, 0.96 GeV], it already contains most of the contribution of the triangle singularity.
In the current calculation, we will just focus on the physics in this phase space range, and such cut will be perfect by reducing the interference from the background.

Additionally, a thin and bright curve in Fig. \ref{fig:loop-dalitz} covers almost $30$ MeV for the invariant mass of $K^+K^-$, which indeed proves our previous argument in Sec. \ref{sec:2-mechanism} that there exists a moving triangle singularity in Fig. \ref{fig:loop-and-background} $(a)$. 
When getting the coordinate of the point on the curve, we can easily find that when the value of the $m_{\pi^+\pi^-}$ is smaller, the $m_{K^+K^-}$ will be larger.
Actually, this phenomena is easy to be understood from the physics of the triangle singularity.
Since the mass of $\psi(2S)$ is fixed, for the $\psi(2S) \to J/\psi \eta$ process, the velocity of the emitted $\eta$ meson is also fixed. However, for the $J/\psi \to \rho^0 \pi^0$ process, when the $m_{\pi^+\pi^-}$ is smaller, it is equivalent to having the mass of the emitted $\rho^0$ smaller, then it will cause the velocity of $\pi^0$ become larger. Obviously, under this situation $\pi^0$ can catch up with $\eta$ easier, which lets the position of triangle singularity in the invariant mass spectrum of $K^+K^-$ become larger.

\begin{figure}[htbp]
  \centering
  \includegraphics[width=1.0\linewidth]{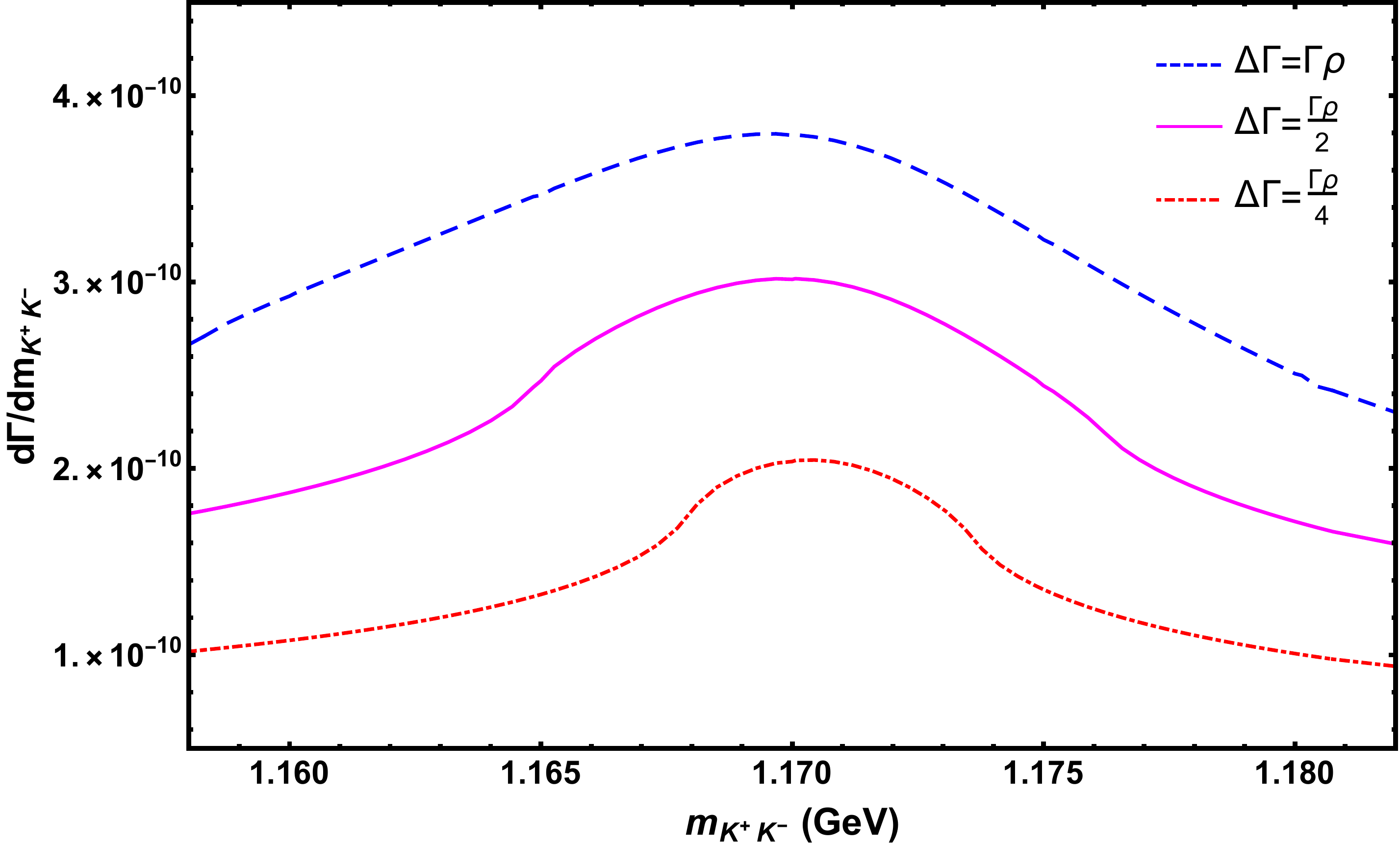}
\caption{The $K^+K^-$ invariant mass spectra of the $\psi(2S) \to \pi^+ \pi^- K^+ K^-$ process, which contains the contribution of loop diagram only. In this figure, the symbol $\Delta\Gamma$ means that the integration on $m_{\pi^+\pi^-}$ is carried out within the interval $[m_\rho-\Delta\Gamma,m_\rho+\Delta\Gamma]$.}
\label{fig:loop-magnitude-1}
\end{figure}

To check the behavior of the triangle singularity in the $K^+K^-$ invariant mass spectrum, an integration on $m_{\pi^+\pi^-}$ is carried out for the intervals as  $[m_\rho-\Gamma_\rho,m_\rho+\Gamma_\rho]$, $[m_\rho-\frac{\Gamma_\rho}{2},m_\rho+\frac{\Gamma_\rho}{2}]$ and $[m_\rho-\frac{\Gamma_\rho}{4},m_\rho+\frac{\Gamma_\rho}{4}]$.
The results are shown in Fig. \ref{fig:loop-magnitude-1}.
For each range of integration, the loop diagram (Fig. \ref{fig:loop-and-background} $(a)$) produces a broad bump in the $K^+K^-$ invariant mass spectrum, and we find that the width of bumps will become narrower when $\Delta\Gamma$ is reduced.
It indicates that this broad width depends on the integration range of $m_{\pi^+\pi^-}$, rather than on the triangle singularity only.
Actually, the peak due to triangle singularity here should be very sharp, because the intermediate states, $J/\psi$, $\eta$ and $\pi^0$ are all very narrow. 
It was proven in our previous work on $\psi(2S) \to p\bar{p}\eta/p\bar{p}\pi^0$ processes \cite{Huang:2020kxf}, where the triangle singularity caused by the pure $J/\psi \eta p$ loop generates a very sharp structure with width around $1$ MeV.

Now let us explain why the integration range of $m_{\pi^+\pi^-}$ affects the width of the bump in the invariant mass spectrum of $K^+K^-$.
As we have mentioned in Sec. \ref{sec:2-mechanism}, the change of $m_{\pi^+\pi^-}$ will cause the movement of the peak position due to the triangle singularity in the $K^+K^-$ invariant mass spectrum. 
In addition, the integration range of $m_{\pi^+ \pi^-}$ is $[m_\rho-\Gamma_\rho,m_\rho+\Gamma_\rho]$, where each value of $m_{\pi^+ \pi^-}$ can have a triangle singularity.
As a result, the broad structure actually is constructed by a series of narrow bumps purely due to the triangle singularities.
Obviously, if we reduce $\Delta\Gamma$, due to the fact that the triangle singularity with $m_{\pi^+\pi^-}$ out of $[m_\rho-\Delta\Gamma,m_\rho+\Delta\Gamma]$ is suppressed tremendously, 
the triangle singularity located around $m_{K^+K^-}=1.17$ GeV, which is the $m_{EF}$ solved from Eq.(\ref{eq:cond1}) with the $m_{CD}=m_{\pi^+\pi^-}=m_{\rho}=0.77$ MeV, 
will give the largest contribution.
At that time, the signal of the triangle singularity will be very sharp, which will give the same behavior as that in $\psi(2S) \to p\bar{p}\eta/p\bar{p}\pi^0$ processes \cite{Huang:2020kxf}.

\begin{figure}[htbp]
  \centering
  \includegraphics[width=1.0\linewidth]{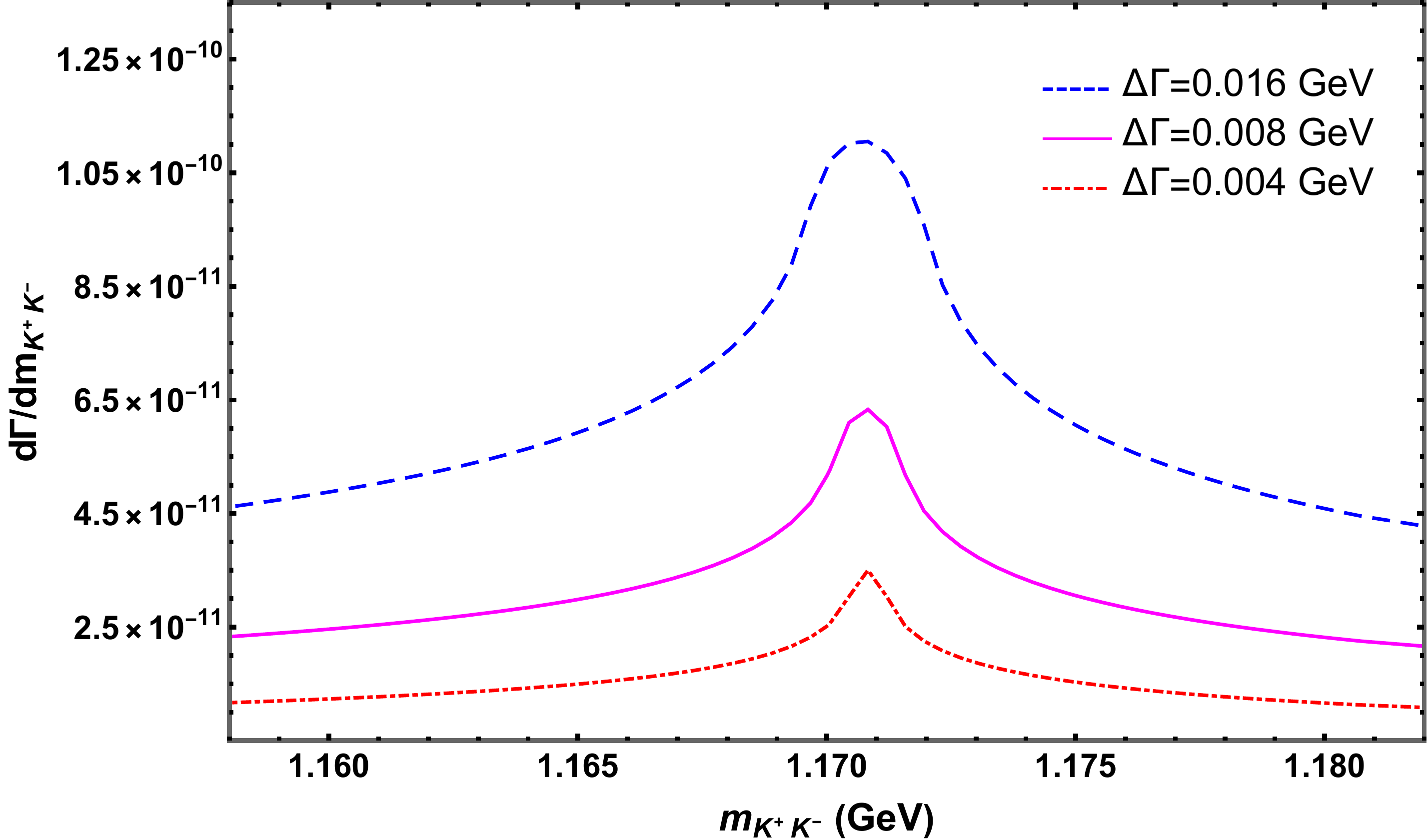}
\caption{The $K^+K^-$ invariant mass spectra of the $\psi(2S) \to \pi^+ \pi^- K^+ K^-$ process, where we contain the contribution of the loop diagram only. In this figure, the symbol $\Delta\Gamma$ means that the integration on $m_{\pi^+\pi^-}$ is carried out within the interval $[m_\rho-\Delta\Gamma,m_\rho+\Delta\Gamma]$.}
\label{fig:loop-magnitude-2}
\end{figure}

In Fig. \ref{fig:loop-magnitude-2}, we continue to reduce $\Delta\Gamma$ to present the $K^+K^-$ invariant mass spectrum of the loop diagram (Fig. \ref{fig:loop-and-background} $(a)$).
The bump caused by triangle singularity becomes much sharper, however, its strength also becomes much smaller.
In summary, a very important information that we can get from the above explanation is that the broad bump in Fig. \ref{fig:loop-magnitude-1} is actually a superposition of a series of sharp triangle singularities with different positions.
Then, through changing the integration range of $m_{\pi^+\pi^-}$, i.e., the value of $\Delta\Gamma$, the width of the peak in the invariant mass spectrum of $K^+K^-$ can be adjusted to a suitable value.
As Ref.\cite{Huang:2020kxf} points out, the width is too narrow to be detected because of the limitation of the resolution of the detector.
Thus, by comparing the structure at all $\Delta\Gamma$ as shown in Figs. \ref{fig:loop-magnitude-1} and \ref{fig:loop-magnitude-2}, we find that $\Delta\Gamma=16$ MeV looks the best choice.

\begin{figure}[htbp]
  \centering
  \includegraphics[width=1.0\linewidth]{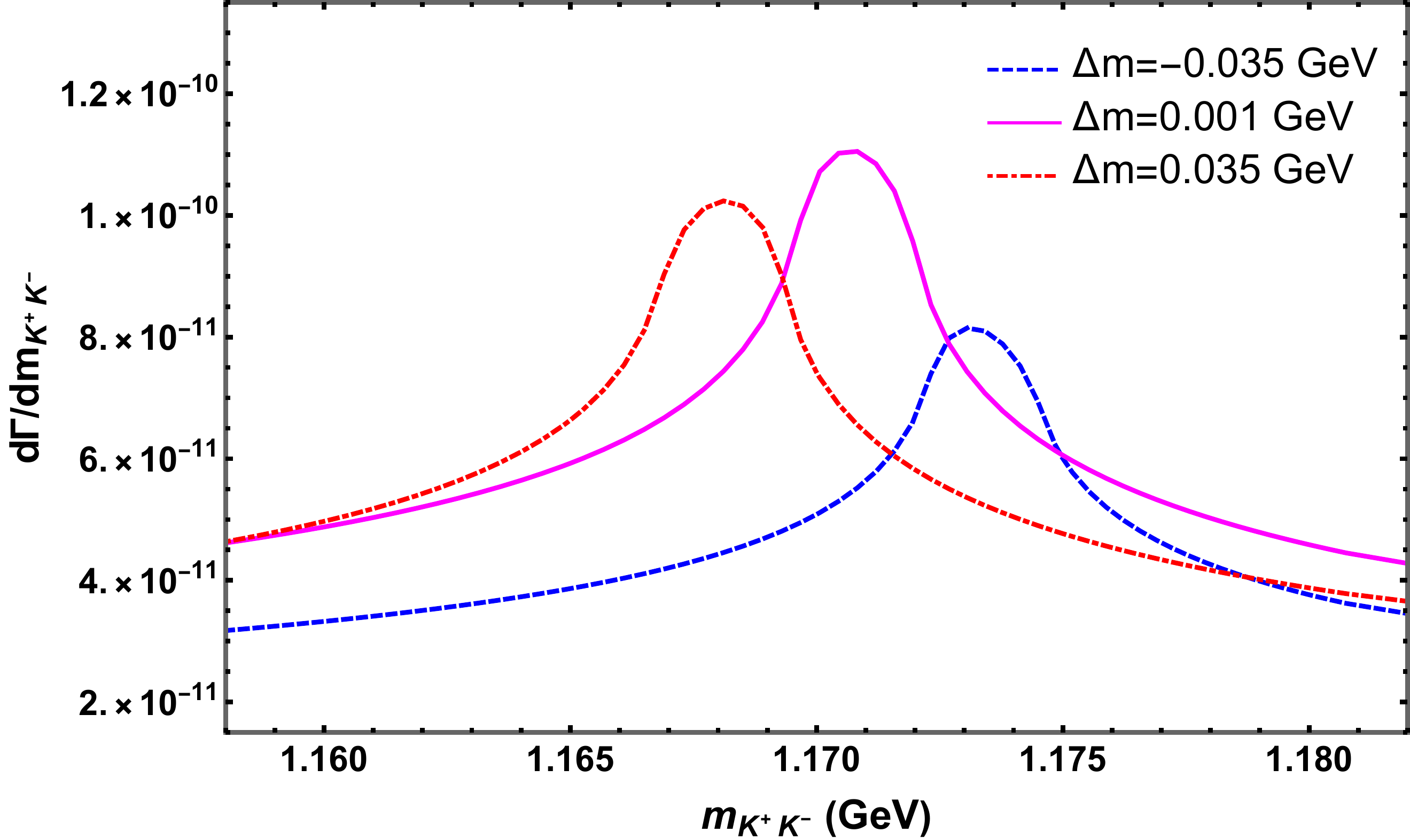}
\caption{The $K^+K^-$ invariant mass spectra of the $\psi(2S) \to \pi^+ \pi^- K^+ K^-$ process, where the $\Delta m$ means that the integration on $m_{\pi^+\pi^-}$ is carried out within the interval $[m_\rho+\Delta m-\Delta\Gamma,m_\rho+\Delta m+\Delta\Gamma]$ with $\Delta\Gamma$ = 16 MeV. In this figure, we only include the contributions of the loop diagram.}
\label{fig:loop-moving}
\end{figure}

Finally, Fig. \ref{fig:loop-moving} presents the movement of the triangle singularity. Here, we carry out the integration on $m_{\pi^+\pi^-}$ within the interval $[m_\rho+\Delta m-\Delta\Gamma,m_\rho+\Delta m+\Delta\Gamma]$, where $\Delta m$ is the shift from the center mass of $\rho$ and $\Delta \Gamma$ is fixed as 16 MeV. 
As shown in this figure, when $\Delta m$ changes, the positions of the peaks caused by the triangle singularities change explicitly.
Furthermore, when $|\Delta m|$ become larger, the strength of the peak is reduced because of the suppression from the propagator of $\rho$ meson. 
For each peak, the width is around $3$ MeV because of the same value of $\Delta \Gamma=16$ MeV used.

\subsection{The possibility of observing this triangle singularity in the $\psi(2S) \to \pi^+ \pi^- K^+ K^-$ process} \label{sec:3-all}

Apparently, to study if our triangle singularity in the $\psi(2S) \to \pi^+ \pi^- K^+ K^-$ process can be observed in experiments, discussions on the loop diagram only is far from enough.
In principle, we should consider the contributions of the tree diagrams given by Fig. \ref{fig:loop-and-background} (b), i.e., the background, to see the possibility.

\begin{figure}[htbp]
  \centering
  \includegraphics[width=1.0\linewidth]{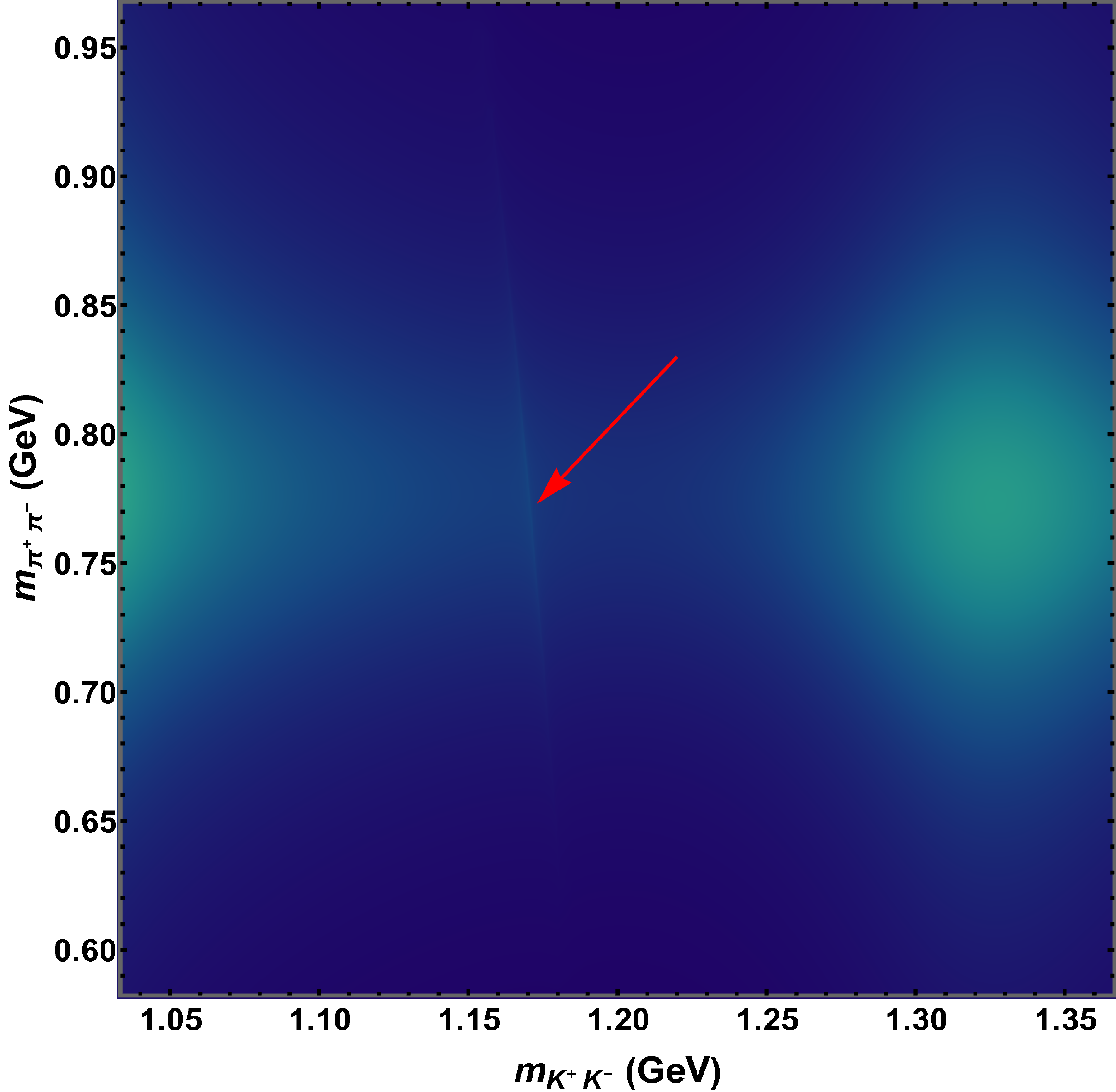}
\caption{The Dalitz plot of the $\psi(2S) \to \pi^+ \pi^- K^+ K^-$ process after considering the contributions of both Fig. \ref{fig:loop-and-background} ($a$) and Fig. \ref{fig:loop-and-background} ($b$), where the intermediate $M$ in Fig. \ref{fig:loop-and-background} ($b$) includes $a_0(980)$ and $a_2(1320)$. Same as Fig. \ref{fig:loop-dalitz}, $m_{\pi^+\pi^-}$ and $m_{K^+K^-}$ are limited in [0.59 GeV, 0.96 GeV] and [1.04 GeV, 1.31 GeV] respectively. In this figure, the two bright spots located around 1 and 1.35 GeV correspond to the contributions of $a_0(980)$ and $a_2(1320)$ respectively. The subtle thin band pointed by the red arrow in the middle is the contribution of the triangle singularity.}
\label{fig:all-dalitz}
\end{figure}

Similarly as done in Sec. \ref{sec:3-Loop}, we also present the Dalitz plot first, and the result is given by Fig. \ref{fig:all-dalitz}. 
In this Dalitz plot, two bright spots located around 1 and 1.35 GeV are for the contributions of $M=a_0(980)$ and $a_2(1320)$ from the tree diagrams as shown in Fig. \ref{fig:loop-and-background} (b), respectively. 
These two bright spots indicate that the contribution of tree diagram dominate in our interested phase space range.
However, fortunately, in Fig. \ref{fig:all-dalitz}, we can still slightly see a subtle thin band at the same location of Fig. \ref{fig:loop-dalitz}, which tells us that the effect caused by our triangle singularity may still be observable.
Thus, next, our task is to find a proper way to make the signal of the triangle singularity still be visible after including the background.

\begin{figure}[htbp]
  \centering
  \includegraphics[width=1.0\linewidth]{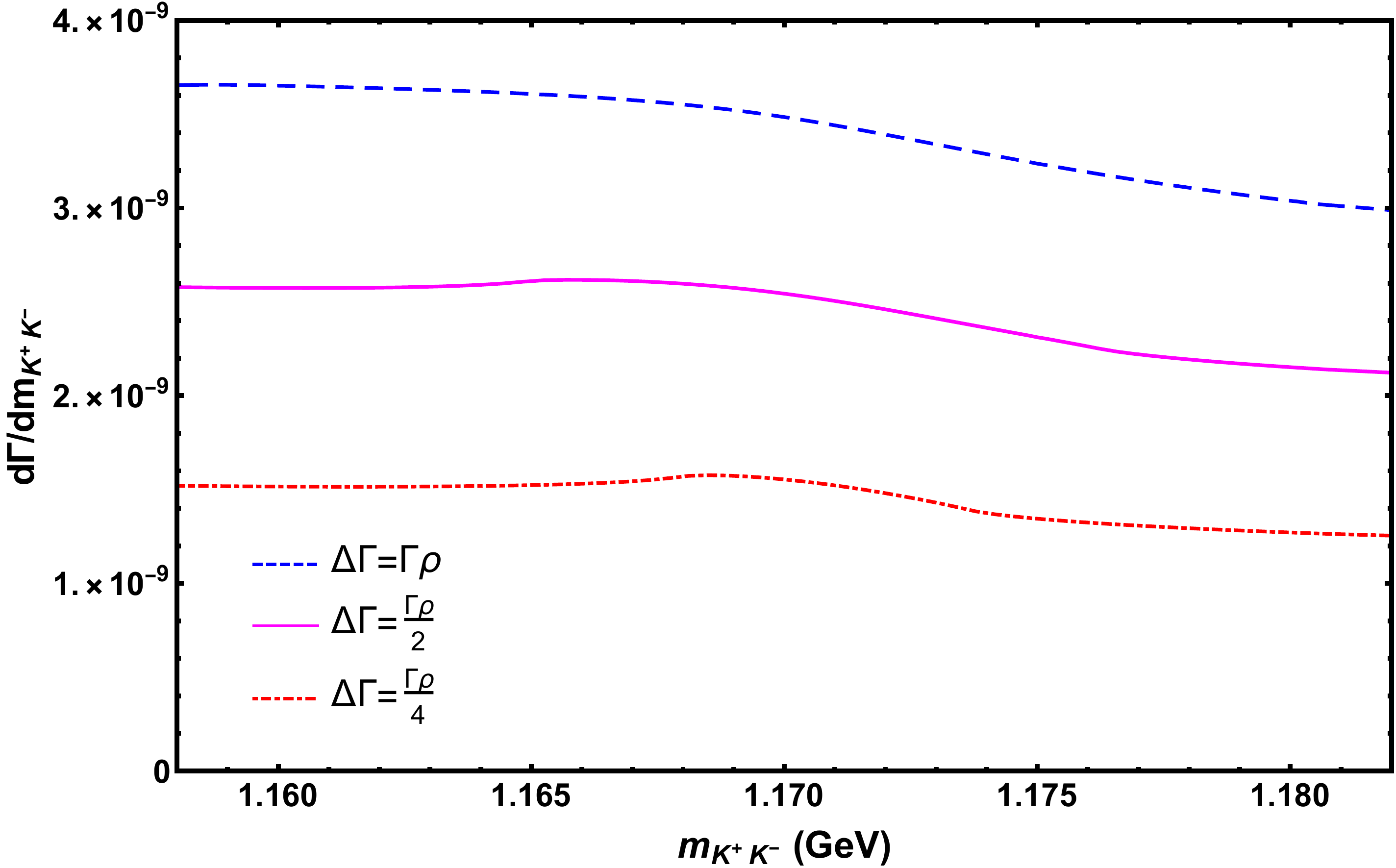}
\caption{The $K^+K^-$ invariant mass spectra of the $\psi(2S) \to \pi^+ \pi^- K^+ K^-$ process, where the $\Delta\Gamma$ means that the integration on $m_{\pi^+\pi^-}$ is carried out within the interval $[m_\rho-\Delta\Gamma,m_\rho+\Delta\Gamma]$. In this figure, both the contributions of the loop diagram and the tree diagrams are considered, where for the tree diagrams, we include the contributions of $a_0(980)$ and $a_2(1320)$.}
\label{fig:loop-all-magnitude}
\end{figure}

\begin{figure}[htbp]
  \centering
  \includegraphics[width=1.0\linewidth]{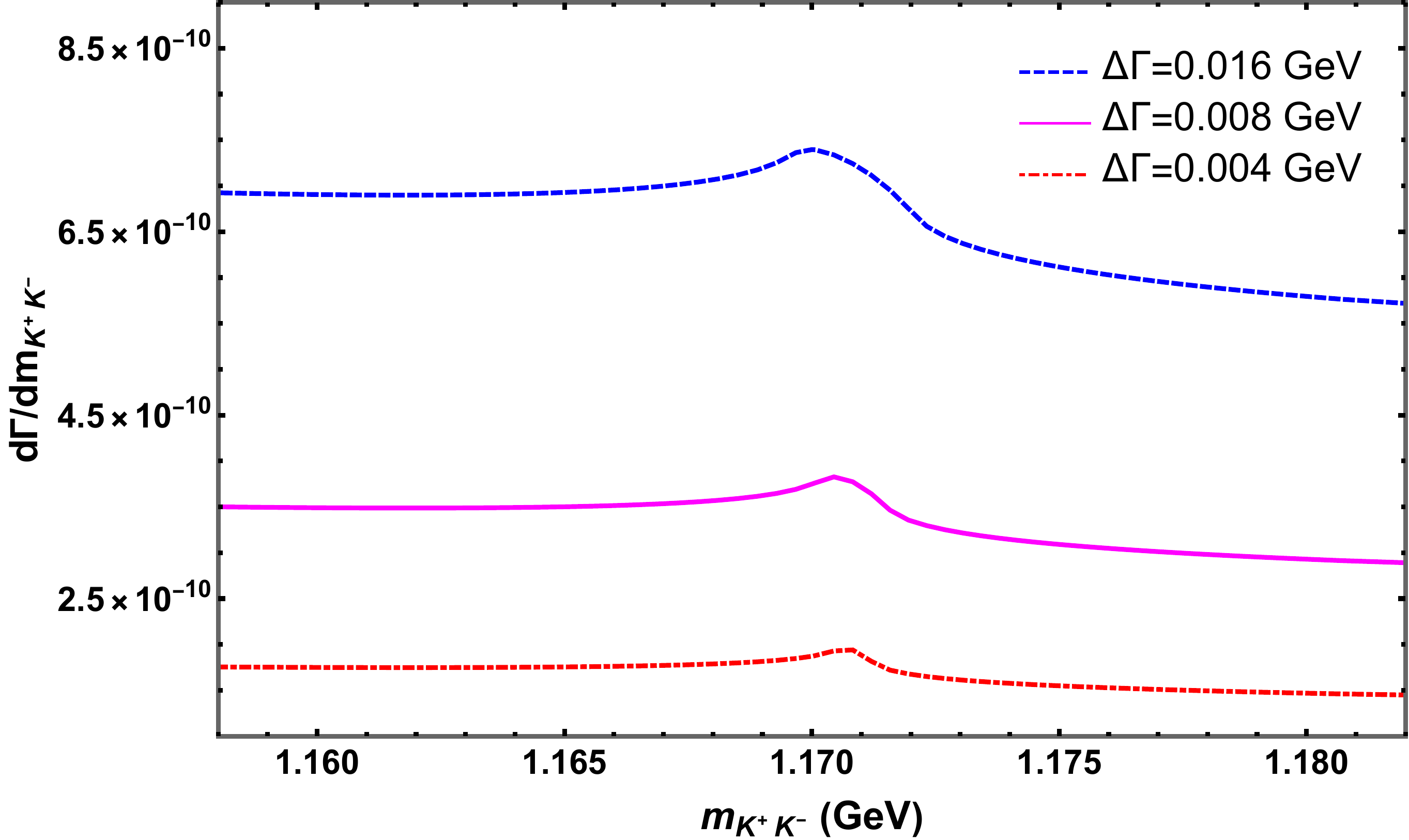}
\caption{The $K^+K^-$ invariant mass spectra of the $\psi(2S) \to \pi^+ \pi^- K^+ K^-$ process, where the $\Delta\Gamma$ means that the integration on $m_{\pi^+\pi^-}$ is carried out within the interval $[m_\rho-\Delta\Gamma,m_\rho+\Delta\Gamma]$. In this figure, the contributions of loop diagram in addition with the contributions of $a_0(980)$ and $a_2(1320)$ are all included.}
\label{fig:ts-magnitude}
\end{figure}

In the first step, we draw the similar figure as Fig.~\ref{fig:loop-magnitude-1} but including the background contribution as shown in Fig.~\ref{fig:loop-all-magnitude}.
Unfortunately, the broad peaks due to the triangle singularity almost disappear and just leave some twists there.
It is natural to continue reducing the value of $\Delta\Gamma$, because the peak structure of the loop becomes much sharper as shown in Fig.\ref{fig:loop-magnitude-2}.
Then we set $\Delta\Gamma$ as 4 MeV, 8 MeV and 16 MeV to see the change of the $K^+K^-$ invariant mass spectrum, and the corresponding numerical results are given in Fig. \ref{fig:ts-magnitude}.
Now the peaks due to the triangle singularity are clearly seen even taking into account the contribution of the tree diagram, which means that it may be possible for experiments to observe our triangle singularity by adding cuts on the invariant mass of $\pi^+\pi^-$.
This phenomenum is understandable.
For each fixed value of $m_{\pi^+\pi^-}$, in the invariant mass spectrum of $K^+K^-$, the triangle singularity only dominates in a very small range. However, the background contribution is everywhere and almost flat.
Then, through the integration, these background contributions are summed together for every point in the invariant mass spectrum of $K^+K^-$.  
Once we reduce the value of $\Delta\Gamma$ roughly with a factor 10, i.e., from $\Delta\Gamma=\Gamma_\rho\sim 140 $ MeV to $16$ MeV, as shown by the dashed blue lines in Fig.\ref{fig:loop-magnitude-1} and Fig.\ref{fig:loop-magnitude-2}, the background contribution is also suppressed roughly with a factor 10, while from Fig.\ref{fig:loop-magnitude-1} and Fig.\ref{fig:loop-magnitude-2} the peak strength of $d\Gamma / dm_{K^+K^-}$ only reduces from $3.8\times 10^{-10}$ to $1.2\times 10^{-10}$.  
In other words, the contribution of background will be suppressed much faster than that of signal when $\Delta\Gamma$ is reduced.
Thus, if the integral interval of $m_{\pi^+\pi^-}$ is too large, the discrepancy between the contributions of tree and loop diagrams will be too large. Furthermore, as discussed before, when $\Delta\Gamma$ is larger, the peak structure of the triangle singularity becomes broader. Thus, the signal of the triangle singularity is buried, which will make our triangle singularity invisible.
Thus, to observe the peak due to the triangle singularity, we need a small interval range of $m_{\pi^+\pi^-}$. 
However, from Fig. \ref{fig:ts-magnitude} we can see that we can not cut $m_{\pi^+\pi^-}$ as small as possible because if the integral interval $m_{\pi^+\pi^-}$ is too small, not only the overall magnitude of the $K^+ K^-$ invariant mass spectrum will be too small, but also the width of the signal will be too narrow, which will make it too difficult for experiments to observe this signal of the triangle singularity. 
Thus, in our view, to detect the triangle singularity in future experiments, we should find a balance between the cut on $m_{\pi^+\pi^-}$, the overall magnitude of $K^+K^-$ invariant mass spectrum and the significance of the signal of the triangle singularity, which we will discuss in detail in Sec. \ref{sec:3-detect}.

\begin{figure}[htbp]
  \centering
  \includegraphics[width=1.0\linewidth]{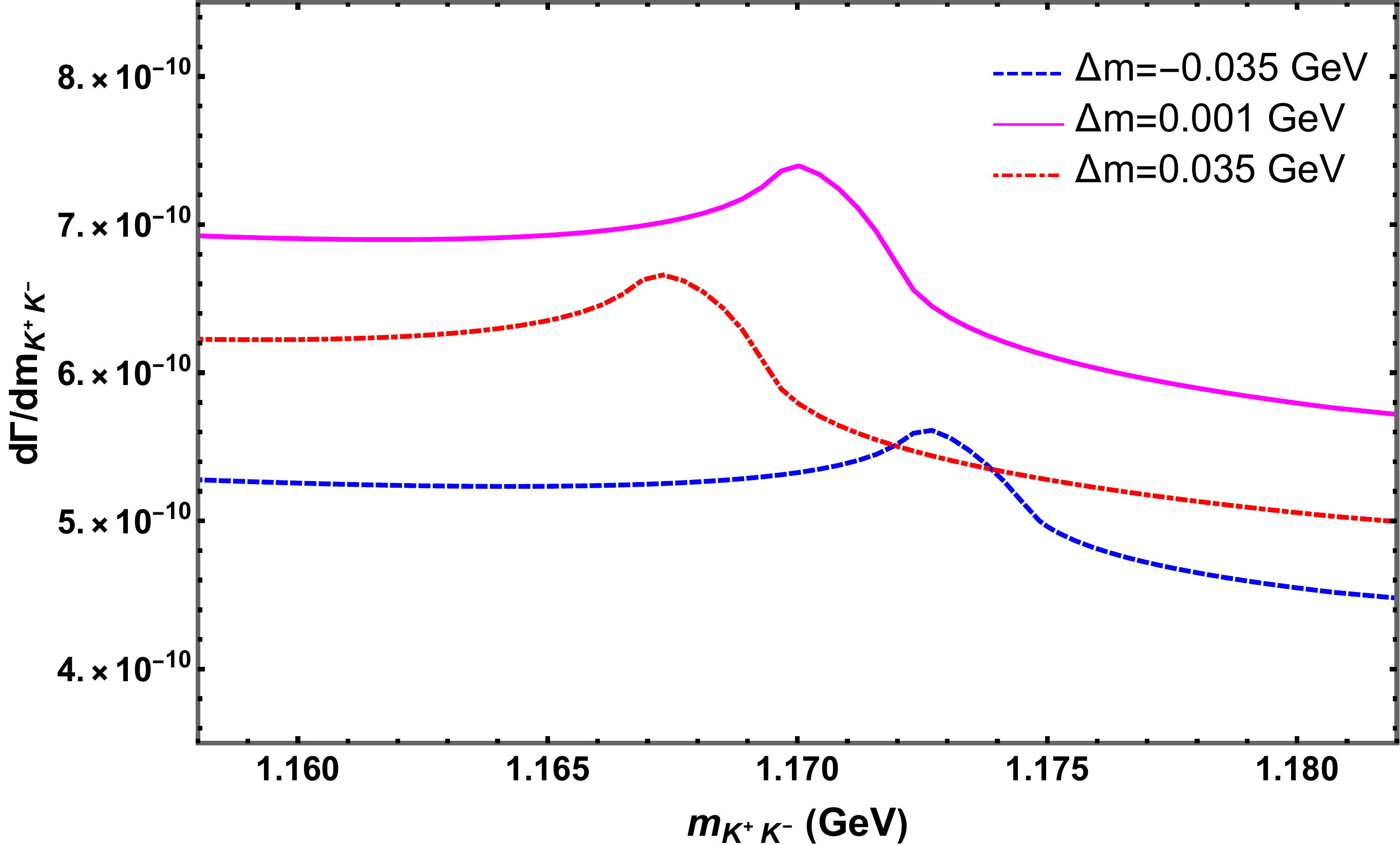}
\caption{The $K^+K^-$ invariant mass spectra of the $\psi(2S) \to \pi^+ \pi^- K^+ K^-$ process, where the $\Delta m$ means that the integration on $m_{\pi^+\pi^-}$ is carried out within the interval $[m_\rho+\Delta m-\Delta\Gamma,m_\rho+\Delta m+\Delta\Gamma]$ with $\Delta\Gamma$ = 16 MeV. In this figure, the contributions of loop diagram in addition with the contributions of $a_0(980)$ and $a_2(1320)$ are all included.}
\label{fig:ts-moving}
\end{figure}

Furthermore, by changing the integral interval of $m_{\pi^+\pi^-}$, Fig. \ref{fig:ts-moving} presents the "moving triangle singularity". 
In Fig. \ref{fig:ts-moving}, the integration on $m_{\pi^+\pi^-}$ is carried out within the interval $[m_\rho+\Delta m-\Delta\Gamma, m_\rho+\Delta m +\Delta\Gamma]$, where $\Delta m$ is the divergence from the center mass of $\rho$ meson and $\Delta\Gamma$ is fixed as 16 MeV. 
From Fig. \ref{fig:ts-moving} we can see that when $|\Delta m| =$35 MeV, the center value of the signal caused by our triangle singularity can move about 3 MeV, in addition, when $\Delta m \neq 0$, both the significance of the signal and the overall magnitude of $K^+K^-$ invariant mass spectrum are decreased, which is easy to be understood since when $\Delta m \neq 0$, both of them will be suppressed by the propagator of $\rho$ meson. 
Thus, 
it requires to change the cuts on $m_{\pi^+\pi^-}$ in experiments to observe the movement of the triangle singularity, which, in our view, must be a very interesting topic and help us understand more about the triangle singularity itself.

At last, it is worthy to make a discussion to justify that it is almost enough to use these tree diagrams instead of all possible background contribution.
We have do other two calculation to estimate the background.
From RPP, we have the branching ratio of the $\psi(2S) \to \pi^+ \pi^- K^+ K^-$ process is $7.2\times 10^{-4}$ \cite{Zyla:2020zbs}.
However, in our calculation the branching ratios of $\psi(2S) \to \rho^0 a^0_0(980) \to \pi^+ \pi^- K^+ K^-$ and $\psi(2S) \to \rho^0 a^0_2(1320) \to \pi^+ \pi^- K^+ K^-$ are around $5 \times 10^{-6}$.
It indicates that both of them are not the dominant processes for the $\psi(2S) \to \pi^+ \pi^- K^+ K^-$ reaction.
From Refs. \cite{BaBar:2007ptr,BaBar:2011btv}, the main contribution is from $\psi(2S) \to K^*(892)\bar{K}^*(892)$.
Thus, we assume the branching ratio of $\psi(2S) \to K^*(892)\bar{K}^*(892)\to \pi^+ \pi^- K^+ K^-$ is $7.2\times 10^{-4}$, i.e., all $\pi^+ \pi^- K^+ K^-$ final states are from $K^*(892)\bar{K}^*(892)$.
But we find that $d\Gamma/dm_{K^+K^-}$ with the integration range of $m_{\pi^+\pi^-}$ from $m_\rho - \Gamma_\rho$ to $m_\rho + \Gamma_\rho$ is around $10^{-11}$, which is an order of magnitude smaller than that of the tree diagrams of $a^0_0(980)$ and $a^0_2(1320)$.
Furthermore, if we just consider a phase space distribution, the $d\Gamma/dm_{K^+K^-}$ by integrating $m_{\pi^+\pi^-}$ from $m_\rho - \Gamma_\rho$ to $m_\rho + \Gamma_\rho$ is around $10^{-10}$ which is still smaller than the contribution of the $a^0_0(980)$ and $a^0_2(1320)$ resonances.
By these two comparisons, we believe that in the range which is sensitive for the signal of the triangle singularity, the main background is roughly from the tree diagram with $a^0_0(980)$ and $a^0_2(1320)$ resonances as calculated above.

\subsection{How to detect the triangle singularity in the $\psi(2S) \to \pi^+ \pi^- K^+ K^-$ process in experiments} \label{sec:3-detect}

Finally, in this subsection we will discuss in detail how to detect the predicted triangle singularity in the $\psi(2S) \to \pi^+ \pi^- K^+ K^-$ process in future experiments. 
As discussed in the above subsection, if the integral interval of $m_{\pi^+\pi^-}$ is too large, the signal of the triangle singularity would be invisible in the invariant mass spectrum of $K^+ K^-$ since the ratio between the contributions of loop and tree diagrams is negligible.
Thus, when analyzing the experimental data of $\psi(2S) \to \pi^+ \pi^- K^+ K^-$ process, experimentalists can not get the signal of the triangle singularity directly from the full $K^+K^-$ invariant mass spectrum, which just integrate the completed range of $m_{\pi^+\pi^-}$. 
Therefore, to extract the triangle singularity, the experimentalists should make cuts on the $m_{\pi^+\pi^-}$ around the $m_{\rho}$ region, and then divide them in several bins of $m_{\pi^+\pi^-}$.

Now let us discuss how to make the cut of $m_{\pi^+\pi^-}$ to show the peak due to triangle singularity on the $m_{K^+K^-}$ spectrum. 
Here, we need to balance two facts, the visible peak structure and enough statistic events, and there exists one variable $\Delta\Gamma$ to control them.
Furthermore, because of the limitation of the resolution of the detector, for example, currently the resolution of BESIII is about 4 MeV \cite{Ablikim:2019hff}, the width of peaks to be observed should not be too small.
Thus, by comparing various lines in Figs.\ref{fig:loop-all-magnitude} and \ref{fig:ts-magnitude}, we think that $\Delta\Gamma$=16 MeV maybe a good choice.
In this situation, the width of the signal is enlarged to about 3 MeV, and the signal almost takes 5\%, which requires almost 500 events at least to control the statistical fluctuations.

Now let us derive the expression to estimate the events number through $d\Gamma/dm_{K^+K^-}$ as follows,
\begin{eqnarray} \label{eq:events}
  N_{\Delta m} = 
  \int_{\Delta m} \frac{d\Gamma}{dm_{K^+K^-}}d m_{K^+K^-} \frac{N_{\psi(2S)}}{\Gamma_{\psi(2S)}} ,\nonumber\\
\end{eqnarray}
where $N_{\psi(2S)}$ is the number of $\psi(2S)$ events, $\Gamma_{\psi(2S)}$ is the width of $\psi(2S)$, $\Delta m$ is the bin width, $N_{\mathrm{\Delta m}}$ is the events that the bin contains, and the integration $\int_{\Delta m} \frac{d\Gamma}{dm_{K^+K^-}}d m_{K^+K^-}$ is the total width under each bin.
Then with Eq. (\ref{eq:events}), as an example, we now analyse if the current BESIII experiment can observe the triangle singularity. 
According to Ref. \cite{Ablikim:2019hff}, nowadays BESIII has 800 million $\psi(2S)$, while in the future, the number of $\psi(2S)$ will increase to 4 billion \cite{Ablikim:2019hff}. 
In addition, currently the resolution of BESIII is about 4 MeV \cite{Ablikim:2019hff}. 
By taking $\Delta\Gamma$=16 MeV, if we have 4 billion $\psi(2S)$, then we will have about 10 events per MeV, also, the width of the signal is enlarged to about 3 MeV.
The enhancement of the signal is about 5\%, which tells it will be very hard for BESIII to observe the triangle singularity.

Recently, we notice that there is a talk on the updated simulation progress of STCF \cite{STCF:talk}. In this talk, we find that in the future STCF, we can get 640 billion $\psi(2S)$ even per year \cite{STCF:talk}, which means even using the $\Delta\Gamma$=16 MeV cut on $m_{\pi^+\pi^-}$, for the $K^+K^-$ invariant mass spectrum, we will have about 1600 events per MeV per year, also, since the enhancement of the signal is about 5\%, we can get 200 events of the enhancement per year. In addition, STCF has excellent resolution \cite{STCF:talk}, which makes us believe STCF will be a very nice platform for searching this triangle singularity.

If the future, experiments can really observe our triangle singularity after doing some cuts on $m_{\pi^+\pi^-}$. Experimentalists are encouraged to change the cuts to verify the movement of our triangle singularity, i.e., experiments can continue to check the results given in Fig. \ref{fig:ts-moving}. In our view, both the detection of our triangle singularity and the observation of its movement are very important and interesting topics. They will not only help us verify the concept of triangle singularity, but also help us understand more about its properties.

\section{Summary}

Although the triangle singularity proposed by L. D. Landau \cite{Landau:1959fi} might be very important in explaining many abnormal experimental results \cite{BESIII:2012aa,Ablikim:2013mio,Liu:2013dau,Xiao:2013iha,Ablikim:2013wzq,Ablikim:2013xfr,
Ablikim:2013emm,Ablikim:2017oaf,Aaij:2015tga,Aaij:2019vzc}. 
The manifestation as a very narrow peak has not yet been seen in experiments because in practical case the particle with mass $m_1$ in Fig.\ref{fig:loop} has a width of the order of tens of MeV.

Thus, Ref. \cite{Huang:2020kxf} predicted precisely that there exists a pure triangle singularity in the $\psi(2S) \to p \bar{p} \eta$ process, which can be observed by the future experiments such as BESIII and STCF. However, apart from searching for the triangle singularity in a process with 3 final states only, we find that it is more interesting to search for a triangle singularity in a process with 4 final states since it will not only give us a triangle singularity, but also this triangle singularity can move in a range.

In the current work, we propose that we can detect such a moving triangle singularity in the $\psi(2S) \to \pi^+ \pi^- K^+ K^-$ process in the invariant mass of $K^+K^-$, we find that the position of the triangle singularity produced by this process can vary from 1.158 to 1.181 GeV.

According to our analysis, it is really possible for future experiments to observe the triangle singularity. However, to detect this triangle singularity and its movement, experimentalists can not observe the $K^+ K^-$ invariant mass spectrum directly, instead, they should do some cuts on the $m_{\pi^+\pi^-}$ first. Based on our numerical results, we suggest that experimentalists can do a series of cuts with about 32 MeV each on the $m_{\pi^+\pi^-}$ around the center mass of $\rho$ meson, i.e., 770 MeV. At that time, the enhancement of the triangle singularity is about 5\%, and the width of the signal is about 3 MeV. In addition, with different cuts on $m_{\pi^+ \pi^-}$, we are confident that experiments can see the movement of the triangle singularity clearly.

Considering the current experimental status, we find that the verification on the triangle singularity in this work may not be observed in the BESIII experiment \cite{Ablikim:2019hff}. However, according to the talk on the updated simulation progress of STCF \cite{STCF:talk}, we find that in the future STCF, we can get about 200 events of the enhancement of our triangle singularity per year, which will be a nice place for detect this triangle singularity. In our view, the observation of this triangle singularity will not only verify the concept of triangle singularity, but also help to understand more about its properties.

\section*{Acknowledgments}

The authors want to thank Rui Chen, Feng-Kun Guo, Satoshi Nakamura, J. A. Oller, Eulogio Oset, and Bing-Song Zou for useful discussions.
This work was supported by the Fundamental Research Funds for the Central Universities,  the Key Research Program of the Chinese Academy of Sciences, Grant NO. XDPB15, and  National
Key R$\&$D Program of China under Contract No. 2020YFA0406400.

\section*{Appendix} \label{sec:2-M}


In Sec. \ref{sec:2-mechanism}, we have written out the amplitudes of triangle loop diagram and tree diagram through effective Lagrangian approach. However, the specific expression of $\mathcal{M}_{\pi^0 \eta \to K^+ K^-}$ in Eq. (\ref{eq:M-Loop}) is still unknown. Thus, in this appendix we will give in detail how we get $\mathcal{M}_{\pi^0 \eta \to K^+ K^-}$.

We adopt the chiral unitary approach for meson meson interaction to describe the $\pi^0 \eta \to K^+ K^-$ scattering. Under this approach, all possible meson meson channels that couple within SU(3) to certain given quantum numbers are considered, then by using the Bethe-Salpeter equation with kernel (potential) taken from chiral Lagrangians in coupled channels \cite{Gasser:1983yg,Bernard:1995dp}, there only remains some regularization scale in the meson meson loop, which can be fitted from the meson meson scattering data. With the chiral unitary apporach, a good agreement with experimental data is obtained up to 1.2 GeV \cite{Oller:1997ti,Oller:1998hw,Kaiser:1998fi,Locher:1997gr,Nieves:1999bx,Pelaez:2006nj}, which exactly covers the range where our triangle singularity moves. Thus, in our view, applying chiral unitary approach to describe the $\pi^0 \eta \to K^+ K^-$ scattering is reasonable.

We start with the leading order chiral Lagrangian, which is expressed as \cite{Gasser:1983yg,Bernard:1995dp,Oller:1997ti,Oller:1998hw,Kaiser:1998fi,Locher:1997gr,Nieves:1999bx,Pelaez:2006nj,Xie:2014tma,Liang:2014tia,Toledo:2020zxj,Ikeno:2021kzf,Molina:2019udw}
\begin{eqnarray}
    \mathcal{L}_2 = \frac{1}{12f^2}\langle (\partial_\mu \Phi \Phi- \Phi \partial_\mu \Phi)^2 + M \Phi^4 \rangle, 
\end{eqnarray}
where
\begin{eqnarray}
    \Phi &=&\begin{pmatrix}
                \frac{\pi^0}{\sqrt{2}}+\frac{\eta}{\sqrt{3}} + \frac{\eta^\prime}{\sqrt{6}} & \pi^+ & K^+\\
                \pi^- & -\frac{\pi^0}{\sqrt{2}}+\frac{\eta}{\sqrt{3}} + \frac{\eta^\prime}{\sqrt{6}} & K^0\\
                K^- & \bar{K}^0 & -\frac{\eta}{\sqrt{3}} + \sqrt{\frac{2}{3}}\eta^\prime
            \end{pmatrix},\nonumber\\
            ~\\
    M&=&\begin{pmatrix}
            m_\pi^2 & 0 & 0\\
            0 & m_\pi^2 & 0\\
            0 & 0 & 2m_K^2-m_\pi^2
        \end{pmatrix}.
\end{eqnarray}

Then, with the leading order chiral Lagrangians above, we can get the leading order $T-\mathrm{matrix}$ elements of the scatterings between $K^+ K^-$,  $K^0 \bar{K}^0$, and $\pi^0 \eta$ as \cite{Xie:2014tma}
\begin{eqnarray}
    (1) && K^+(k) K^-(p) \to K^+(k^\prime) K^-(p^\prime)\nonumber\\
        &&\qquad t_1 = -\frac{s}{2f^2},\\
    (2) && K^0(k) \bar{K}^0(p) \to K^0(k^\prime) \bar{K}^0(p^\prime)\nonumber\\
        &&\qquad t_2=t_1,\\
    (3) && K^+(k) K^-(p) \to K^0(k^\prime) \bar{K}^0(p^\prime)\nonumber\\
        &&\qquad t_3 = \frac{1}{2}t_1,\\
    (4) && K^+(k) K^-(p) \to \pi^0(k^\prime) \eta(p^\prime)\nonumber\\
        &&\qquad t_4 = -\frac{\sqrt{3}}{12f^2}(3s-\frac{1}{3}m_\pi^2-\frac{8}{3}m_K^2-m_\eta^2),\\
    (5) && K^0(k) \bar{K}^0(p) \to \pi^0(k^\prime) \eta(p^\prime)\nonumber\\
        &&\qquad t_5 = -t_4,\\
    (6) && \pi^0(k) \eta(p) \to \pi^0(k^\prime) \eta(p^\prime)\nonumber\\
        &&\qquad t_6 = -\frac{m_\pi^2}{3f^2},
\end{eqnarray}
where $s = (k+p)^2$ and $f=0.093$ GeV \cite{Xie:2014tma}.

Next, after considering the isospin relation, we can get the following $T^{(2)}-\mathrm{matrix}$ elements as \cite{Xie:2014tma}
\begin{eqnarray}
    T^{(2)}_{K\bar{K} \to K\bar{K}} &=& \frac{1}{2}(t_1 - t_3 - t_3 + t_2)\nonumber\\
    &=& -\frac{s}{4f^2},\\
    T^{(2)}_{K\bar{K} \to \pi\eta}&=&T^{(2)}_{\pi\eta \to K\bar{K}}=T^{(2)}_{K\bar{K} \to \pi^0\eta} = -\frac{1}{\sqrt{2}}(t_4-t_5)\nonumber\\
    &=& \frac{\sqrt{6}}{12f^2}(3s-\frac{1}{3}m_\pi^2-\frac{8}{3}m_K^2-m_\eta^2),\\
    T^{(2)}_{\pi\eta \to \pi\eta} &=& T^{(2)}_{\pi^0\eta \to \pi^0\eta} = t_6\nonumber\\
    &=& -\frac{m_\pi^2}{3f^2}.
\end{eqnarray}

Thus, the leading order coupled channel $T-\mathrm{matrix}$, i.e., $T_2$, is
\begin{eqnarray}
    T_2 =   \begin{pmatrix}
                T^{(2)}_{\pi\eta \to \pi\eta} & T^{(2)}_{\pi\eta \to K\bar{K}}\\
                T^{(2)}_{K\bar{K} \to \pi\eta} & T^{(2)}_{K\bar{K} \to K\bar{K}}
            \end{pmatrix}.
\end{eqnarray}

Then, the final $T-\mathrm{matrix}$ can be written as
\begin{eqnarray}
    T =   \begin{pmatrix}
      T_{\pi\eta \to \pi\eta} & T_{\pi\eta \to K\bar{K}}\\
      T_{K\bar{K} \to \pi\eta} & T_{K\bar{K} \to K\bar{K}}
  \end{pmatrix} = [1-T_2 G]^{-1} T_2,
\end{eqnarray}
where $G$ is a diagonal matrix of loop functions, whose diagonal matrix elements can be written as
\begin{eqnarray}
    G(s) &=& i \int \frac{d^4 q}{(2 \pi)^4} \;
    \frac{1}{q^2 - M_1^2 + i \epsilon} \;
    \frac{1}{ (P - q)^2 - M_2^2 + i \epsilon}\nonumber\\
    & = & \frac{1}{32 \pi^2}\Bigg[-\frac{\Delta}{s}\log\frac{M_1^2}{M_2^2}+2\frac{\Delta}{s}\log\frac{1+\sqrt{1+\frac{M_1^2}{q_{\mathrm{max}}^2}}}{
      1+\sqrt{1+\frac{M_2^2}{q_{\mathrm{max}}^2}}}\nonumber\\
    &&-2\log \left\{ \left(1+\sqrt{1+
    \frac{M_1^2}{q_{\mathrm{max}}^2}}\right)\left(1+\sqrt{1+\frac{M_2^2}{q_{\mathrm{max}}^2}}\right) 
    \right\}\nonumber\\
    &&+\log\frac{M_1^2 \, M_2^2}{q_{\mathrm{max}}^4}+\frac{\nu}{s}
    \left\{ \log\frac{s-\Delta+\nu\sqrt{1+\frac{M_1^2}{q_{\mathrm{max}}^2}}}
    {-s+\Delta+\nu \sqrt{1+\frac{M_1^2}{q_{\mathrm{max}}^2}}} \right.\nonumber\\
    &&+ \left.\log\frac{s+\Delta+\nu\sqrt{1+\frac{M_2^2}{q_{\mathrm{max}}^2}}}
    {-s-\Delta+\nu \sqrt{1+\frac{M_2^2}{q_{\mathrm{max}}^2}}} \right\}  \Bigg],
\end{eqnarray}
with $\nu=\sqrt{(s-(M_1+M_2)^2)(s-(M_1-M_2)^2)}$, $\Delta=M_1^2-M_2^2$, and $q_{\mathrm{max}}$ is the cutoff, which is set as $q_\mathrm{max} = 0.6$ GeV \cite{Xie:2014tma,Liang:2014tia,Toledo:2020zxj,Ikeno:2021kzf,Molina:2019udw}.

Finally, considering the isospin relation, the amplitude of $\pi^0\eta \to K^+ K^-$ process can be expressed by the $T$-matrix element as
\begin{eqnarray}
  \mathcal{M}_{\pi^0\eta \to K^+ K^-} = \frac{1}{2} T_{\pi\eta \to K \bar{K}}.
\end{eqnarray}

\vfil

\end{document}